\documentclass[aps,pra,reprint,superscriptaddress,nofootinbib]{revtex4-2}

\usepackage{bm}
\usepackage{amsmath,amssymb}
\usepackage{mathtools}
\usepackage{array}
\usepackage{booktabs}
\usepackage{multirow}
\usepackage{graphicx}
\usepackage[colorlinks=true,citecolor=blue,linkcolor=magenta]{hyperref}

\newcommand{\ket}[1]{|#1\rangle}
\newcommand{\bra}[1]{\left\langle#1\right|}

\newcommand{\Tr}{\operatorname{Tr}}

\newcommand{\spec}{\operatorname{spec}}

\newcommand{\pstab}{\mathfrak p}

\begin{document}

\title{Symmetry-resolved fermionic correlations in iron-oxide ground spaces}

\author{Federico Petrovich}
\affiliation{Information Sciences, Los Alamos National Laboratory, Los Alamos, New Mexico 87545, USA}

\author{Daniel Staros}
\affiliation{Theoretical Division, Los Alamos National Laboratory, Los Alamos, New Mexico 87545, USA}

\author{Daniel N. Blaschke}
\affiliation{X-Computational Physics Division, Los Alamos National Laboratory, Los Alamos, New Mexico 87545, USA}

\author{Jian-Xin Zhu}
\affiliation{Theoretical Division, Los Alamos National Laboratory, Los Alamos, New Mexico 87545, USA}
\affiliation{Center for Integrated Nanotechnology, Los Alamos National Laboratory, Los Alamos, New Mexico  87545, USA}

\author{M. Cerezo}
\thanks{cerezo@lanl.gov}
\affiliation{Information Sciences, Los Alamos National Laboratory, Los Alamos, New Mexico 87545, USA}
\thanks{cerezo@lanl.gov}

\begin{abstract}
Symmetries of quantum systems provide a natural way to organize the eigenspaces of many-body Hamiltonians. When degeneracies exist,  however, they play a more subtle role as different sets of symmetry operators can resolve the same energy subspace into states with very different correlation properties. We study this interplay using a sequence of FeO and Fe$_2$O$_2$ models. For a Hubbard model of FeO, the ground space is 16-fold degenerate. Resolving it by local magnetizations gives a basis of Slater determinants, while total-spin quantum numbers select entangled and non-Gaussian states from the same ground manifold. Adding density-density Hund interactions breaks the independent spin rotations and reduces the ground space to a spin-reversal pair with maximal magnetization. These states are extremely close to Slater determinants in their natural occupations, yet have finite two-body cumulants and conserved-operator algebras whose orbit dimensions differ substantially from the Slater values. Fe$_2$O$_2$ provides a larger example in which spatial mode entanglement can be substantial while fermionic non-Gaussianity remains weak, with the conserved-operator structure again distinguishing the state from the corresponding Slater transformation class. Together, these examples show that symmetry resolution, mode entanglement, reduced-density-matrix correlations, and conserved-operator structure provide complementary information about interacting and degenerate electronic ground states.
\end{abstract}

\maketitle

\section{Introduction}

Quantum information is widely recognized as an extremely powerful tool for analyzing, characterizing, and classifying the correlation properties of ground states of many-body Hamiltonians~\cite{amico2008entanglement,eisert2010colloquium,de2018genuine,laflorencie2016quantum,tarabunga2024critical,sarkis2025molecules,deside2025detecting,moca2025non,tarabunga2025efficient,jasser2025stabilizer,cianciulli2024bipartite,gigena2021many,collura2024quantum,gu2025magic}. For a nondegenerate Hamiltonian, it is extremely natural to simply compute different correlation properties of its ground state. When the ground space is degenerate, however, the situation becomes a bit more ambiguous. Here, it is entirely possible that different states within that subspace can have very different correlation properties. This issue is well known and has been extensively reported in the study of degenerate and symmetry-broken many-body systems, where one uses the symmetry information of the Hamiltonian to select and resolve particular states within the degenerate manifold~\cite{koma1994symmetry,okatev2025exploring,cerezo2015nontransverse,cerezo2017factorization,petrovich2022ground}. Specifically, one projects onto the simultaneous eigenspaces of commuting symmetry operators of the model.

For fermionic systems, even after a particular state has been selected (say from a degenerate manifold) one must still decide which type of correlations to study. Here one must acknowledge that there is a strong historical bias in quantum information toward correlations defined between distinguishable subsystems. This is natural from the information-processing perspective in which much of entanglement theory was developed~\cite{wilde2013quantum}, where one typically imagines Alice and Bob in spatially separated laboratories and asks which correlations can be manipulated using local operations and classical communication~\cite{horodecki2009quantum}. Applied to fermionic systems, this viewpoint leads one to partition the fermionic modes into distinguishable sets and study bipartite mode entanglement. This is a perfectly meaningful notion of correlation, but there is no a priori reason for it to capture all correlations relevant to indistinguishable fermions.

Over the past few decades, however, there has been growing interest in correlations that are genuinely fermionic, in the sense that they go beyond those imposed by antisymmetrization. For pure fixed-particle-number states, this amounts to studying deviations from the uncorrelated Slater determinants (SDs)~\cite{schliemann2001quantum,gigena2015entanglement,gigena2017bipartite,di2019fermionic,gigena2020one,gigena2021many,cianciulli2024bipartite,benatti2020entanglement}. Indeed, departures from the SD manifold reveal intrinsically many-body fermionic structure, including fractional natural occupations, failures of Wick factorization, and transformation properties that cannot occur for a single determinant. Recent work has developed several complementary ways of probing this structure through reduced density matrices, fermionic non-Gaussianity, property testing, resource theories, and invariant or symmetry-based characterizations~\cite{garcia2026fermionic,bittel2025optimal,tarabunga2026computable,braccia2026commutant,sierant2026theory,west2026fermionic,petrovich2026symmetry}.

In this work, we study the implications of different choices of correlation in a sequence of low-energy FeO and Fe$_2$O$_2$ models. We focus on these systems as iron oxides provide prototypical strongly correlated materials that have long served as benchmarks for electronic-structure methods. Moreover, they are also a natural target for quantum simulation, where low-energy models derived from electronic-structure calculations provide a route toward studying correlated materials on quantum hardware  (see our companion paper \cite{staros2026towards}). We consider first a Hubbard model of FeO, then a Hubbard-Hund model with the same orbital structure, and finally a larger Fe$_2$O$_2$ model. For each case, we determine the symmetries and ground-space structure and then study the resulting states using mode entanglement, reduced-density-matrix measures of fermionic correlation, and the conserved-operator structure associated with invertible one-particle transformations.

\begin{figure}
\centering
\includegraphics[width=1\linewidth]{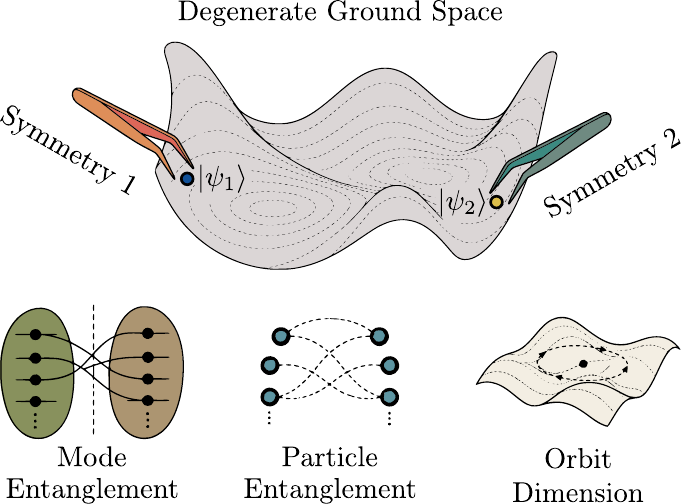}
\caption{\textbf{Schematic representation of our results.} Different sets of commuting symmetry operators can select different states, here $\ket{\psi_1}$ and $\ket{\psi_2}$, from the same degenerate ground space manifold. We characterize the resulting states from three perspectives: Mode entanglement probes correlations across a chosen partition of the fermionic modes, particle entanglement probes correlations beyond the SD description, and the dimension of the fermionic orbit characterizes the state under invertible one-particle transformations.}
\label{fig:schematic}
\end{figure}

In the Hubbard FeO model, different symmetry resolutions of the same sixteen-dimensional ground space select states with qualitatively different correlation properties, ranging from exact SDs to entangled and non-Gaussian states. Adding Hund interactions removes most of this degeneracy and leaves a pair of maximally polarized ground states whose natural occupations remain extremely close to those of a SD, even though their conserved-operator algebras place them on substantially different fermionic orbits. In Fe$_2$O$_2$, sizable entanglement across spatial partitions coexists with weak fermionic non-Gaussianity. As schematically shown in Fig.~\ref{fig:schematic}, the correlation structure assigned to these ground states therefore depends both on how the degenerate manifold is resolved and on which notion of correlation is being considered.

\section{The $\rm FeO$ Hamiltonians and their symmetries}
\label{sec:models}

In this section, we introduce two Hamiltonians for the FeO molecule, a prototypical strongly correlated system. We begin with a Hubbard model and then incorporate density-density Hund interactions, keeping the same orbital structure and one-body terms. Both Hamiltonians are constructed within a low-energy description based on DFT and projected Wannier functions for quantum simulation. For each Hamiltonian, we determine its spatial, particle-number, and spin symmetries and use them to resolve the structure of its ground space before studying its correlation properties. We refer the reader to our companion paper \cite{staros2026towards} for additional details on the construction of these Hamiltonians.

The Fe--O bond is taken to lie along the $x$ axis and the model retains the $5$ Fe $3d$ orbitals $\{d_{xy}, d_{yz}, d_{z^2}, d_{xz}, d_{x^2-y^2}\}$ together with the $3$ O $2p$ orbitals $\{p_x, p_y, p_z\}$, giving a total of $n=16$ spin orbitals. The choice of the $x$ axis along the bond is useful because the orbital hybridization then separates naturally according to the symmetry of the orbitals with respect to the bond. Specifically, one can first consider the spatial reflection symmetries along the $y$  and $z$ axes, which we respectively denote as $P_y$, $P_z$.  Indeed, these allow us to classify the eight spatial orbitals
according to their parities $(\pi_y,\pi_z)$, with $\pi_y,\pi_z=\pm1$, into four groups
\begin{equation}
\begin{aligned}
\mathcal C_{\scriptscriptstyle ++} &= \{p_x,d_{z^2},d_{x^2-y^2}\}, & \mathcal C_{\scriptscriptstyle +-} &= \{p_z,d_{xz}\},\\
\mathcal C_{\scriptscriptstyle -+} &= \{p_y,d_{xy}\},
& \mathcal C_{\scriptscriptstyle --} &= \{d_{yz}\}.
\end{aligned}
\label{eq:FeOgroups}
\end{equation}

In what follows we will find it convenient to decompose the electronic FeO Hamiltonian into two terms as
\begin{equation}
H_{\rm FeO}=h_{\rm FeO}+V_{\rm FeO},
\end{equation}
where $h_{\rm FeO}$ contains the one-body hopping terms and $V$ is the electron-electron Coulomb interaction. Then, define as $a_{rs}^{\dagger}$ and $a_{rs}$ the fermionic creation and annihilation operators, respectively, for an electron in spatial orbital $r\in \mathcal C_{\pm\pm}$ with spin $s\in\{\uparrow,\downarrow\}$. 
The one-body Hamiltonian $h_{\rm FeO}$ must preserve the spatial reflection symmetries of the molecule and therefore cannot couple orbitals with different parities. Hence, we can write it as
\begin{equation}
h_{\rm FeO}=\sum_{\pi_y,\pi_z=\pm}h_{\pi_y\pi_z},
\label{eq:FeOonebody}
\end{equation}
where
\begin{equation}
\quad h_{\pi_y\pi_z} = \sum_{s=\uparrow,\downarrow}
\sum_{r,r'\in\mathcal C_{\pi_y\pi_z}}
h^{\pi_y\pi_z}_{rr'}a_{rs}^\dagger a_{r's}.
\label{hpiypiz}
\end{equation}
The explicit matrices ${\bf h}^{\pi_y\pi_z}$ are given in Appendix~\ref{app:FeO}.

The decomposition \eqref{eq:FeOonebody}, \eqref{hpiypiz}, implies that the one-body
Hamiltonian has independent particle-number and spin symmetries in each orbital group. Thus, the associated particle-number and spin operators are
\begin{align}
\hat{N}_{\pi_y\pi_z}
&=\sum_{r\in\mathcal C_{\pi_y\pi_z}}
\sum_{s=\uparrow,\downarrow} n_{rs},\\
S_{\pi_y\pi_z}^\mu
&=\frac12\sum_{r\in\mathcal C_{\pi_y\pi_z}}
\sum_{s,s'} \sigma^\mu_{ss'}a^\dagger_{rs} a_{rs'},
\qquad \mu=x,y,z,
\end{align}
with $n_{rs} = a^\dagger_{rs}a_{rs}$ and ${\bm \sigma}^\mu$ the Pauli matrices, we have
\begin{equation}
[h_{\rm FeO},\hat{N}_{\pi_y\pi_z}]=0,
\quad
[h_{\rm FeO},S_{\pi_y\pi_z}^\mu]=0,\label{eqs:one-body-sims}
\end{equation}
leading to an independent $U(1)\times SU(2)$ symmetry for each of the four $(\pi_y,\pi_z)$ orbital groups.

Moving on, the Coulomb interaction $V_{\rm FeO}$ breaks the independent particle-number and spin symmetries of the orbital groups $(\pi_y,\pi_z)$. However, it must still preserve the global particle-number and spin symmetries, whose operators are
\begin{subequations}\label{eqs:total-ops}
\begin{eqnarray}
\hat{N} &=& \sum_{\pi_y,\pi_z = \pm} \hat{N}_{\pi_y\pi_z}, \\
S^\mu &=& \sum_{\pi_y,\pi_z = \pm} S^\mu_{\pi_y\pi_z},
\end{eqnarray}    
\end{subequations}
as well as the spatial reflections 
\begin{equation}
P_y=e^{i\pi(\hat N_{\scriptscriptstyle -+}+\hat N_{\scriptscriptstyle --})},
\qquad
P_z=e^{i\pi(\hat N_{\scriptscriptstyle +-}+\hat N_{\scriptscriptstyle --})}.
\end{equation}

In the next sections, we present two approximations to the Coulomb interaction $V$, examining which symmetries of the one-body and the full electronic Hamiltonian are preserved or broken in each case. The resulting FeO Hamiltonians are
\begin{equation}
H_{\rm FeO}^{h}=h_{\rm FeO}+V_{\rm FeO}^{h},\qquad H_{\rm FeO}^{hh}=h_{\rm FeO}+V_{\rm FeO}^{hh},
\end{equation}
where $V_{\rm FeO}^{h}$ and $V_{\rm FeO}^{hh}$ denote the Hubbard and Hubbard--Hund interactions, respectively.  We restrict both models to the sector with $N=12$ electrons.

\subsection{Hubbard model of FeO}
\label{subsec:hubfeo}

In the first FeO Hamiltonian considered, the interaction acts only on the five Fe $d$ orbitals and is taken to be the on-site Hubbard interaction
\begin{equation}
V_{\rm FeO}^{h} = U \sum_{r\in\mathcal B_d} n_{r\uparrow}n_{r\downarrow},
\label{eq:V1}
\end{equation}
with $U=4.0~{\rm eV}$. We again refer the reader to \cite{staros2026towards} for a detailed justification of these models. This interaction preserves all the independent particle-number and spin symmetries of the one-body Hamiltonian:
\begin{equation}
[V_{\rm FeO}^{h},\hat{N}_{\pi_y\pi_z}]=0,
\quad
[V_{\rm FeO}^{h},S_{\pi_y\pi_z}^\mu]=0,
\label{eq:H1sym}
\end{equation}
for all $\pi_y,\pi_z = \pm$, $\mu=x,y,z$.

Combining Eqs.~\eqref{eqs:one-body-sims} and~\eqref{eq:H1sym} we see that both terms in $H^{h}$ commute with the four particle-number and spin symmetries, meaning that the full Hamiltonian $H^{h}$ also preserves these symmetries. Consequently, the $12$ operators $\{\hat N_{\pi_y\pi_z}, {\bm S}_{\pi_y\pi_z}^2, S_{\pi_y\pi_z}^z\}_{\pi_y\pi_z}$ form a complete set of mutually commuting symmetries whose corresponding eigenvalues define the quantum numbers $(N_{\pi_y\pi_z}, S_{\pi_y\pi_z}, M_{\pi_y\pi_z})$ allowing us to diagonalize $H^{h}$ in each independent symmetry block.

Moreover, since the Hubbard interaction acts independently within each orbital group, the full Hamiltonian separates as
\begin{equation}
H_{\rm FeO}^{h}= \sum_{\pi_y,\pi_z=\pm}H_{\pi_y\pi_z}^{h},
\quad [H_{\pi_y\pi_z}^{h},
H_{\pi_y'\pi_z'}^{h}]=0,
\label{eq:Hfactor}
\end{equation}
and then, each $H_{\pi_y\pi_z}^{h}$ can be diagonalized independently. Each contribution commutes with the corresponding local spin operators $\mathbf S_{\pi_y\pi_z}^2$ and $S_{\pi_y\pi_z}^z$, allowing a further block diagonalization. Consequently, the
original $N=12$ problem of dimension $\binom{16}{12}=1820$ reduces to independent symmetry blocks of dimension at most $8$. In addition, the independent local spin-rotation symmetries imply that each energy level of $H^{h}$ is at least $g=\prod_{\pi_y,\pi_z}
(2S_{\pi_y\pi_z}+1)$ -fold degenerate, corresponding to the possible values of the local magnetizations $M_{\pi_y\pi_z}=-S_{\pi_y\pi_z},\ldots,
S_{\pi_y\pi_z}$.

{\it The ground space.} We now turn our attention to the ground space of $H_{\rm FeO}^{h}$. Comparing the lowest energies of the symmetry blocks, we find that the ground space belongs to the sector
\begin{equation}
N_{++} = 5, \quad N_{-+} = N_{+-} = 3, \quad N_{--} = 1.
\label{eq:GSsectoroccupations}
\end{equation}

Equation~\eqref{eq:GSsectoroccupations} has a simple interpretation. The four orbital groups $\mathcal C_{\pi_y \pi_z}$ contain respectively 3, 2, 2, and 1 spatial orbitals, and can therefore accommodate at most 6, 4, 4, and 2 electrons. Thus, this result shows that each group is missing exactly one electron from complete filling. Indeed, defining the hole-number operators as
\begin{equation}
\hat{N}_{\pi_y \pi_z}^h=2|\mathcal C_{\pi_y \pi_z}|-\hat{N}_{\pi_y \pi_z},
\end{equation}
we obtain
\begin{equation}
N^h_{\pi_y \pi_z} = 1 \ \forall \ \pi_y, \pi_z = \pm.
\label{eq:oneholepergroup}
\end{equation}

To determine the ground-state energy, we introduce the particle-hole representation ${\bar a}_{rs}^\dagger=a_{rs}$. Within the sector of Eq.~\eqref{eq:oneholepergroup}, the two-body Hubbard interaction between holes vanishes identically, and the Hamiltonian reduces to
\begin{equation}
\left.H_{\rm FeO}^{h}\right|_{\rm GS}=\bar E^{h}\mathbb I +\bar h_{\rm FeO},
\end{equation}
where 
\begin{equation}
\bar E^{h}= 2\sum_{\pi_y,\pi_z=\pm}
\rm{Tr}({\bf h}^{\pi_y\pi_z})+5U  =-82.6~{\rm eV}
\end{equation}
is the energy of the completely filled configuration and $\bar h_{\rm FeO}$ is a one-body Hamiltonian for the holes, obtained from the matrices ${\bf h}^{\pi_y\pi_z}$ by reversing the sign of all matrix elements and subtracting $U$ from the diagonal entries corresponding to Fe $d$ orbitals.

Since the $4$ orbital groups are disconnected, we can express the one-body hole Hamiltonian as 
\begin{equation}
    \bar h_{\mathrm{FeO}}=\sum_{\pi_y,\pi_z=\pm} \bar h_{\pi_y\pi_z}^{h}\,.
\end{equation}
Thus, denoting by $\lambda_{\pi_y\pi_z}$ the lowest eigenvalue of each $\bar h_{\pi_y\pi_z}^{h}$, we simply obtain
\begin{equation}
E_{\rm GS}^{h} =
\bar E^{h}+\sum_{\pi_y,\pi_z = \pm}^4\lambda_{\pi_y\pi_z} =
-77.8456836238~{\rm eV}.
\label{eq:E1gs}
\end{equation}
That is, the ground state energy is obtained by adding to the completely filled configuration the individual lowest energies in each term.  Moreover, we find that the lowest one-hole eigenvalue $\lambda_{\pi_y\pi_z}$ of all orbital groups corresponds to $S_{\pi_y\pi_z}=1/2$, then being twofold degenerate with respect to $M_{\pi_y\pi_z}=\pm1/2$. Hence, each group contributes an effective spin-$1/2$ degree of freedom. This implies that the full Hamiltonian has a degenerate ground space of 
\begin{equation}
\mathcal G^{h}
\simeq
(\mathbb C^2)^{\otimes4},
\qquad
\dim\mathcal G^{h}=2^4=16.
\label{eq:GSstructure}
\end{equation}

This degeneracy can be further resolved by diagonalizing the four local magnetizations $S_{\pi_y\pi_z}^z$. Specifically, diagonalizing each $\bar h_{\pi_y\pi_z}$ in the sectors $M_{\pi_y\pi_z}=\pm1/2$, it is spanned by
\begin{equation}
\ket{M_{\scriptscriptstyle ++},M_{\scriptscriptstyle +-}, M_{\scriptscriptstyle -+},M_{\scriptscriptstyle --}} \!=\! \prod_{\pi_y,\pi_z=\pm}
\bar a_{\scriptscriptstyle \pi_y\pi_z,M_{\pi_y\pi_z}}^{\dagger} \ket{\bar 0},
\label{eq:localGSdet}
\end{equation}
with $M_{\pi_y\pi_z}\!=\!\pm\frac12$, where $\ket{\bar 0}$ is the completely filled electron state and $\bar a_{\pi_y\pi_z,M_{\pi_y\pi_z}}^{\dagger}$ creates a hole in the lowest orbital of group $\mathcal C_{\pi_y\pi_z}$ with spin projection $M_{\pi_y\pi_z}$. Notably, each of these states is a SD. 

At this point we encounter the first limit of a symmetry-resolved description of the ground space. This is due to the fact that some of the symmetry generators of the Hamiltonian do not commute with one another, and hence the ground state degeneracy does not admit a unique symmetry resolution. Indeed, different choices of commuting observables can therefore select different states from the same degenerate manifold. In particular, we recall that besides the four local magnetizations $S_{\pi_y\pi_z}^z$, the Hamiltonian also commutes with the global $z$-spin operator $S^z$ as defined in Eq.~\eqref{eqs:total-ops} and with $\mathbf S^2$, where we define  
\begin{eqnarray}
\mathbf S^2= (S^x)^2 + (S^y)^2 + (S^z)^2  \,.
\end{eqnarray}
However, the local magnetizations do not commute with the total spin, as in general,
\begin{equation}
[S_{\pi_y\pi_z}^z,\mathbf S^2]\neq0\,.
\end{equation}
Thus, diagonalizing the local magnetizations and diagonalizing total-spin operators lead to different bases of the same degenerate ground space.

As such, when resolving the ground space as per the eigenvalues of total spin, coupling the $4$ effective spin-$1/2$ degrees of freedom yields
\begin{equation}
\left(\frac12\right)^{\otimes4} = 2[0]\oplus3[1]\oplus[2],
\label{eq:spinDecomp}
\end{equation}
corresponding to $2$ singlets, $3$ triplets,
and $1$ quintet, implying that the dimensions of the total-magnetization sectors are $(1,4,6,4,1)$ for $M = -2,-1,0,1,2$ respectively. These two symmetry resolutions select states with different fermionic correlation properties, as discussed below.

\subsection{Hubbard-Hund interaction in FeO}
\label{subsec:hundfeo}

We next consider the second FeO Hamiltonian, which includes the Hund interaction. The latter  reads
\begin{align}
V_{\rm FeO}^{hh}
={}& U\sum_{r\in\mathcal B_d}n_{r\uparrow}n_{r\downarrow} + U'\sum_{\substack{r<r'\\
r,r'\in\mathcal B_d}}\left( n_{r\uparrow}n_{r'\downarrow} + n_{r\downarrow}n_{r'\uparrow} \right) \nonumber\\
& +(U'-J_H) \sum_{\substack{r<r'\\r,r'\in\mathcal B_d}} \left(n_{r\uparrow}n_{r'\uparrow} + n_{r\downarrow}n_{r'\downarrow}\right),
\label{eq:V2}
\end{align}
with $\mathcal B_d=\{d_{xy}, d_{yz}, d_{z^2}, d_{xz}, d_{x^2-y^2}\}$ and 
\begin{equation}
U=4.5~{\rm eV}, \quad U'=2.7~{\rm eV}, \quad J_H=0.9~{\rm eV}.
\label{V2par}
\end{equation}
This interaction retains the density-density terms of the Hubbard--Kanamori model, omitting spin-flip exchange and pair hopping, and modifies the symmetry structure of the Hubbard model in two important ways. 

We next analyze how the additional interaction changes the symmetries found for the Hubbard model. The spatial symmetries are unaffected, meaning that
\begin{equation}
[H_{\rm FeO}^{hh},P_y]=[H_{\rm FeO}^{hh},P_z]=0,
\end{equation}
and that the $\mathbb Z_2\times\mathbb Z_2$ reflection symmetry and the four parity sectors discussed above remain unchanged. On the one hand, $V_{\rm FeO}^{hh}$ breaks the independent local spin-rotation symmetries, so that neither the local nor the total squared-spin operators are generally conserved. That is,
\begin{equation}
[H_{\rm FeO}^{hh},{\bm S}_{\pi_y \pi_z}^2]\neq0, \quad
[H_{\rm FeO}^{hh},{\bm S}^2]\neq0.
\end{equation}
Although the latter symmetry is preserved by
the full Coulomb Hamiltonian, it is broken by the density-density approximation itself. 

Despite the previous, the eight operators $\{N_{\pi_y \pi_z},S_{\pi_y \pi_z}^z\}_{\pi_y \pi_z}$ remain mutually commuting symmetries of $H_{\rm FeO}^{hh}$, generating a continuous $U(1)^8$ symmetry that allows us to block-diagonalize $H_{\rm FeO}^{hh}$ into sectors labeled by $(N_{\pi_y \pi_z},M_{\pi_y \pi_z})_{\pi_y,\pi_z}$. In addition, the Hamiltonian is also invariant under global spin reversal. Indeed, define as $F$ the operator exchanging spin-up and spin-down on every orbital, i.e., 
\begin{eqnarray}
    F S_{\pi_y \pi_z}^z F^\dagger=-S_{\pi_y \pi_z}^z\,,
\end{eqnarray}
then
\begin{equation}
    [F, H_{\rm FeO}^{hh}] = 0\,.
\end{equation}
The importance of this symmetry is that it directly ensures that every energy level with at least one nonzero $M_{\pi_y \pi_z}$ is at least twofold degenerate.

Finally, we note that the density-density interactions couple different orbital groups energetically. As such, $H^{hh}$ can no longer be decomposed into four independent Hamiltonians as in Eq. \eqref{eq:Hfactor}. 

{\it The ground space.} As in the Hubbard model, we find that the ground space belongs to the sector
\begin{equation}
N_{++} = 5, \quad N_{-+} = N_{+-} = 3, \quad N_{--} = 1.
\end{equation}
However, the Hund interaction in \eqref{eq:V2} now couples the orbital groups, so that the effective Hamiltonian remains interacting, unlike in the previous model. Nevertheless, the hole in $\mathcal C_{--}=\{d_{yz}\}$ remains localized, since this group contains a single isolated orbital and then, only the $3$ remaining holes can exhibit nontrivial fermionic correlations.

The ground space is now twofold degenerate, and is further restricted to the two sectors $M_{\pi_y \pi_z}=\pm1/2$ for all $\pi_y, \pi_z=\pm$. This corresponds to total magnetizations of $M=\pm2$. Singling out the $M=+2$ state, the ground state takes the form
\begin{equation}
\ket{\Psi_+^{{\rm FeO}}}=
\ket{1_{8\uparrow}}
\otimes \ket{\psi_{4\downarrow}},
\label{eq:Hundfactor}
\end{equation}
where $\ket{1_{8\uparrow}}$ denotes the completely filled spin-up sector, and $\ket{\psi_{4\downarrow}}$ describes the $4$ correlated spin-down electrons in
$\mathcal C_{++},\mathcal C_{+-},\mathcal C_{-+}$. The fermionic correlations of $\ket{\psi_\downarrow^{(3)}}$ (which is not a SD) are analyzed below. The $M=-2$ ground state follows by global spin reversal and explains the twofold degeneracy.

Finally, combining the conserved occupation operators allows us to reduce the interacting
ground-state problem to a block of dimension $3\times2\times2=12$, whose exact diagonalization yields 
\begin{equation}
E_{\rm GS}^{hh}
=-47.763926823962~{\rm eV}.
\label{eq:E2gs}
\end{equation}

\section{The $\rm Fe_2O_2$ Hamiltonian and its symmetries}
\label{subsec:fe2o2}

We now consider the larger ${\rm Fe}_2{\rm O}_2$ model, which contains $2$-Fe and $2$-O centers. Spatially, the $4$ atoms are arranged in a square (atoms of the same element occupy opposite corners of the square). As in the FeO model, we retain the five $3d$ orbitals of each Fe atom and the three $2p$ orbitals of each O atom. Denoting the two centers of each element by $i=1,2$, we define the corresponding orbital sets $\{d_{xy}^{(i)},d_{yz}^{(i)},d_{z^2}^{(i)}, d_{xz}^{(i)},d_{x^2-y^2}^{(i)}\}$, and $\{p_x^{(i)},p_y^{(i)},p_z^{(i)}\}$. The model therefore contains $16$ spatial orbitals, or equivalently $n=32$ spin orbitals.

Unlike the FeO molecule, the Fe$_2$O$_2$ square geometry exhibits the three independent spatial reflection symmetries, $P_x$, $P_y$, $P_z$, allowing us to classify the orbitals according to their parities $(\pi_x,\pi_y,\pi_z)$, with $\pi_x,\pi_y,\pi_z=\pm1$ as before. To this end, we introduce the symmetric and antisymmetric combinations
\begin{align}
p_\mu &= \tfrac{p_\mu^{(1)}+p_\mu^{(2)}}{\sqrt{2}},
& \Delta p_\mu &= \tfrac{p_\mu^{(1)}-p_\mu^{(2)}}{\sqrt{2}},\\
d_\alpha &= \tfrac{d_\alpha^{(1)}+d_\alpha^{(2)}}{\sqrt{2}},
& \Delta d_\alpha &= \tfrac{d_\alpha^{(1)}-d_\alpha^{(2)}}{\sqrt{2}},
\end{align}
for $\mu = x,y,z$ and $\alpha = xy, yz, xz, x^2-y^2, z^2$. Using this notation, the $16$ spatial orbitals then separate into $8$ groups
\begin{equation}
\hspace{-0.5em}
\begin{aligned}
\mathcal C_{\scriptscriptstyle +++} &= \{d_{z^2},d_{x^2-y^2},\Delta p_y\},
& \mathcal C_{\scriptscriptstyle ++-} &= \{\Delta d_{xz},p_z\},\\
\mathcal C_{\scriptscriptstyle +-+} &= \{\Delta d_{xy},p_y\},
& \mathcal C_{\scriptscriptstyle +--} &= \{d_{yz},\Delta p_z\},\\
\mathcal C_{\scriptscriptstyle -++} &= \{\Delta d_{z^2},\Delta d_{x^2-y^2},p_x\},
& \mathcal C_{\scriptscriptstyle -+-} &= \{d_{xz}\},\\
\mathcal C_{\scriptscriptstyle --+} &= \{d_{xy},\Delta p_x\},
& \mathcal C_{\scriptscriptstyle ---} &= \{\Delta d_{yz}\}.
\end{aligned}
\label{eq:Fe2O2groups}
\end{equation}

Using the same fermionic notation as above, the full electronic Hamiltonian reads
\begin{equation}
H_{\rm{Fe_2O_2}}=h_{\rm{Fe_2O_2}}+V_{\rm{Fe_2O_2}},
\end{equation}
where $h_{\rm{Fe_2O_2}}$ contains the one-body hopping terms and $V_{\rm{Fe_2O_2}}$ is again the Coulomb interaction. As before, the molecular geometry requires the one-body Hamiltonian to preserve the spatial reflection symmetries $P_x$, $P_y$, $P_z$, and hence prevents couplings between orbitals with different parities. As such, we can decompose the operator $h_{\rm{Fe_2O_2}}$ as
\begin{equation}
h_{\rm Fe_2 O_2}=\sum_{\pi_x,\pi_y,\pi_z=\pm}
h_{\pi_x\pi_y\pi_z},
\label{eq:Fe2O2onebody}
\end{equation}
where
\begin{equation}
h_{\pi_x\pi_y\pi_z} =
\sum_{s=\uparrow,\downarrow}
\sum_{r,r'\in\mathcal C_{\pi_x\pi_y\pi_z}}
h_{rr'}^{\pi_x\pi_y\pi_z}
a_{rs}^\dagger a_{r's}.
\label{eq:Fe2O2blocks}
\end{equation}
The explicit one-body matrices are given in
Appendix~\ref{app:fe2ham}.

As in the FeO model, the decomposition above implies independent particle-number and spin
symmetries for each orbital group, whose respective operators are given by
\begin{align}
\hat{N}_{\pi_x\pi_y\pi_z}
&=\sum_{r\in\mathcal C_{\pi_x\pi_y\pi_z}}
\sum_{s=\uparrow,\downarrow} n_{rs},\\
S_{\pi_x\pi_y\pi_z}^\mu
&=\frac12\sum_{r\in\mathcal C_{\pi_x\pi_y\pi_z}}
\sum_{s,s'} \sigma^\mu_{ss'}a^\dagger_{rs} a_{rs'}.
\end{align}
Importantly, the Coulomb interaction $V_{\rm{Fe_2O_2}}$ breaks the independent orbital symmetries, while preserving the global particle-number and $SU(2)$ spin symmetries, as well as the three spatial reflections $P_x$, $P_y$, and $P_z$.

In the following, we investigate which of the one-body symmetries are preserved by the density-density Hubbard--Hund approximation to the Coulomb interaction, and use the resulting symmetry structure to block-diagonalize the Hamiltonian and determine its ground space. We will restrict this model to the sector with $N=24$ electrons. 

\subsection*{Hubbard-Hund interaction in $\rm Fe_2O_2$}

The Hubbard-Hund interaction $V^{hh}_{\rm Fe_2O_2}$ has the same form as that in Eq. \eqref{eq:V2}, with the same parameters. The only difference is that now it contains the $10$ Fe $3d$ orbitals of both centers.

On the one hand, although this interaction does not generally preserve the particle numbers as before, it preserves their corresponding occupation parities, given by
\begin{equation}
P_{\pi_x\pi_y\pi_z}
=e^{i\pi\hat N_{\pi_x\pi_y\pi_z}}.
\label{eq:Fe2O2parities}
\end{equation}
Moreover, the interaction does preserve the particle-number and the $z$-component spin
operator associated with the two larger orbital groups of opposite parity under $P_z$. That is, defining the set of orbitals
\begin{equation}
\mathcal C_\pm=
\bigcup_{\pi_x,\pi_y=\pm}
\mathcal C_{\pi_x\pi_y\pm},
\end{equation}
the corresponding conserved operators are
\begin{align}
\hat{N}_{\pm}
&=\sum_{r\in\mathcal C_{\pm}}
\sum_{s=\uparrow,\downarrow} n_{rs},\\
S_{\pm}^\mu
&=\frac12\sum_{r\in\mathcal C_{\pm}}
\sum_{s,s'} \sigma^\mu_{ss'}a^\dagger_{rs} a_{rs'}.
\end{align}
Thus, we finally have
\begin{equation}
[V^{hh}_{\rm Fe_2O_2},P_{\pi_x\pi_y\pi_z}] \!=\!
[V^{hh}_{\rm Fe_2O_2},\hat N_\pm] \!=\!
[V^{hh}_{\rm Fe_2O_2},S_\pm^z]=0,
\end{equation}
and since the one-body Hamiltonian also preserves
all these symmetries, they remain conserved by the full Hubbard--Hund Hamiltonian. That is
\begin{equation}
\hspace{-0.13cm}
[H^{hh}_{\rm Fe_2O_2},P_{\pi_x\pi_y\pi_z}] \!=\!
[H^{hh}_{\rm Fe_2O_2},\hat N_\pm] \!=\!
[H^{hh}_{\rm Fe_2O_2},S_\pm^z]\!=\!0.
\end{equation}

At this point, we find it important to note that only $6$ of the $8$ occupation parities are independent once the conserved particle numbers $\hat N_\pm$ are fixed. This is due to the fact that
\begin{equation}
\prod_{\pi_x,\pi_y=\pm} P_{\pi_x\pi_y\pm} = e^{i\pi\hat N_\pm},
\end{equation}
implying that, within a sector of fixed $N_+$ and $N_-$, the product of the four occupation parities associated with each sector is fixed. Consequently, only three of the four parities in each sector can be chosen independently, with the fourth determined by the other three and the corresponding particle number. Thus, together with the four conserved quantities $\hat N_\pm$ and $\hat S_\pm^z$, we obtain a total of $10$ independent symmetries. As before, the density-density interaction breaks the spin rotation symmetry which directly implies that 
\begin{eqnarray}
[H_{\rm Fe_2O_2}^{hh},\mathbf S^2]\neq0\,.
\end{eqnarray}
Still, global spin reversal remains an exact symmetry, implying that every energy level with at least one nonzero $M_+$ or $M_{-}$ is at least twofold degenerate.

Combining all the previous results we know that the $N=24$ electron Hilbert space has dimension $\binom{32}{24}=10\,518\,300$ and decomposes into symmetry blocks. Enumerating all occupation configurations compatible with the $4$ conserved quantum numbers $(N_+,M_+,N_-,M_-)$ and the $6$ independent occupation parities yields $7292$ nonempty symmetry blocks. The largest block has dimension $18\,400$ and corresponds to 
\begin{equation}
    (N_+,M_+,N_-,M_-)=(14,0,10,0)\,,
\end{equation}
with all eight occupation parities equal to $+1$. Thus, the identified symmetries reduce the original Hamiltonian to blocks that are more than $500$ times smaller in dimension, substantially simplifying the exact diagonalization problem.

{\it The ground space}. Exact diagonalization of the symmetry-resolved Hamiltonian gives
\begin{equation}
E_{\rm GS}^{\rm Fe_2O_2}=-10.839851578034~{\rm eV},
\label{eq:Fe2E}
\end{equation}
with a GS pair obtained by global spin reversal $(M_+,M_-) \to (-M_+,-M_-)$. The positive-polarization GS lies in the sector 
\begin{equation}
    (N_+,N_-)=(15,9)\,,\qquad (M_+,M_-)=(5/2,3/2)\,,
\end{equation}
or equivalently
\begin{equation}
    (N_{+\uparrow},N_{-\uparrow})=(10,6)\,,\qquad (N_{+\downarrow},N_{-\downarrow})=(5,3)
\end{equation}
with parities
\begin{equation}
    P_{\pm+\mp}\!=\!1, \qquad P_{\pm+\pm}=P_{\pm-\pm}=P_{\pm-\mp}=-1\,,
\end{equation}
having dimension $80$. Since it has all $16$ spin-up orbitals occupied, it is convenient to introduce the particle-hole representation as in the FeO model. Thus, the ground state problem then reduces to $8$ spin-down holes, with occupations $(N_{+\downarrow}^h,N_{-\downarrow}^h)=(5,3)$. 

In this sector, the effective two-body interaction depends only on the total number of
spin-down holes in the Fe $3d$ orbitals and therefore commutes with the individual occupation operators $\hat N_{\pi_x\pi_y\pi_z}$. This implies that the effective Hamiltonian recovers the $8$ independent particle-number symmetries, allowing a further decomposition of the ground state parity sector into two independent blocks of dimensions
$72$ and $8$. 

The ground state lies in the $72$-dimensional block characterized by
\begin{equation}
\hspace{-0.35cm}
\begin{aligned}
N_{\scriptscriptstyle +++}\!=\!5,\!\!\quad
N_{\scriptscriptstyle \pm + \mp }\!=\!4,\!\!\quad
N_{\scriptscriptstyle +-\pm}
\!=\!N_{\scriptscriptstyle --+}\!=\!3, \!\!\quad N_{\scriptscriptstyle - \pm - } \!=\!1
\end{aligned}
\label{Nsector}
\end{equation}
or equivalently, in the hole representation,
\begin{equation}
\hspace{-0.1cm}
N^h_{\scriptscriptstyle ++-}\!=\!0,\quad
N^h_{\scriptscriptstyle -++}\!=\!2,\quad
N^h_{\scriptscriptstyle \pm\pm\pm}
\!=\!N^h_{\scriptscriptstyle +-\pm}
\!=\!N^h_{\scriptscriptstyle -\pm\mp}\!=\!1.
\label{Nhsector}
\end{equation}
These occupations imply that the two spin-down orbitals in $\mathcal C_{++-}$ are fully occupied, while those in $\mathcal C_{-+-}$ and $\mathcal C_{---}$ are exactly empty. Consequently, only $12$ of the $16$ spatial orbitals remain active, and the positive-polarization ground state has the form
\begin{equation}
|\Psi^{\rm Fe_2O_2}_+\rangle
=|1_{16\uparrow}\rangle
\otimes|\psi_{6\downarrow}\rangle
\otimes|1_{2\downarrow}\rangle,\label{eq:fe2o2factor}
\end{equation}
where $|1_{16\uparrow}\rangle$ denotes the
completely filled spin-up sector, $|\psi_{6\downarrow} \rangle$ describes the $6$ correlated spin-down electrons in the $12$ remaining active orbitals, and $|1_{2\downarrow}\rangle$ denotes the $2$ fully occupied spin-down orbitals in $\mathcal C_{++-}$.

\section{Mode entanglement and fermionic correlations}
\label{sec:standardent}

When discussing correlations in a fermionic many-body state, it is important to first specify what notion of correlation is being considered. For instance, one can import the usual notion of entanglement which was developed for systems of distinguishable constituents. There, a bipartition of the Hilbert space $\mathcal H=\mathcal H_A\otimes\mathcal H_B$ specifies two subsystems, and a pure state is unentangled when it factorizes as $\ket{\Psi}=\ket{\psi_A}\otimes\ket{\psi_B}$. Entanglement then captures quantum correlations between these distinguishable subsystems. 

For identical fermions, however, the individual particles cannot be used in this way as distinguishable subsystems, meaning that we cannot simply ask if a given fermion is ``entangled'' with another one. Hence, if one is keen on using notions such as bipartite entanglement, one instead partitions the fermionic modes into two distinguishable sets. This is known as mode entanglement. In the models considered here, natural choices include spin-up versus spin-down modes, Fe versus O orbitals, or the orbitals associated with one atomic center versus the rest of the molecule. Once such a partition is chosen, the corresponding Fock spaces define the subsystems and standard bipartite entanglement measures can be applied.

This construction is useful, but it also makes clear what such entanglement is measuring. It quantifies correlations between the chosen sets of modes and is therefore tied to that particular partition. This is not the same question as asking whether the fermionic many-body state contains correlations beyond those already present in a single SD. Indeed, this has led the community to also define the so-called particle entanglement, or genuine fermionic correlations. These characterize genuine correlations arising in fermionic systems beyond those imposed by antisymmetrization. Importantly, these do not require a particular partition of the modes. As we show below, these two notions capture different aspects of the same quantum states and need not exhibit the same behavior.

\subsection{Mode entanglement}

The starting point of mode entanglement is the identification of fermionic Fock states with occupation-number states,
\begin{equation}
a_{i_1}^\dagger\cdots a_{i_N}^\dagger|0\rangle
\longleftrightarrow
|n_1,\ldots,n_n\rangle,
\end{equation}
where $n_i=1$ if $i\in\{i_1,\ldots,i_N\}$ and $n_i=0$ otherwise, with a fixed ordering of the fermionic modes. This allows us to treat each fermionic mode as a two-level subsystem and apply the usual notion of bipartite entanglement to a partition of the modes into two sets $A$ and $\bar A$. The corresponding Fock spaces define the subsystems, and a pure state is unentangled across this partition when it factorizes as $|\Psi\rangle=|\psi_A\rangle\otimes|\psi_{\bar A}\rangle$.

In general, the entanglement across this partition is characterized by the eigenvalues of the reduced density matrix
\begin{equation}
{\bm \rho}_A=\mathrm{Tr}_{\bar A}(|\Psi\rangle\langle\Psi|),
\end{equation}
which coincide with the nonzero eigenvalues of $\rho_{\bar A}$ for a pure global state.
A standard measure is the von Neumann entropy ~\cite{nielsen2000quantum},
\begin{equation}
S_A=-\mathrm{Tr}({\bm \rho}_A\log_2{\bm \rho}_A).
\end{equation}

{\it Spatial mode entanglement.}
We first consider the local-magnetization ground states of the Hubbard FeO Hamiltonian $H_{\rm FeO}^{h}$. As previously shown, the ground space can be spanned by the $16$ states \eqref{eq:localGSdet} which are exact SD. Nevertheless, they all exhibit the same non-zero entanglement entropy across the Fe--O mode partition,
\begin{equation}
S^{h}_{\mathrm{Fe}|\mathrm O}=0.860032.
\end{equation}
This finite entropy arises because the occupied one-particle orbitals are hybridized combinations of Fe and O orbitals, independently of their spin orientations. Thus, a SD can exhibit finite mode entanglement, reflecting orbital delocalization across the chosen partition rather than fermionic correlations beyond antisymmetrization.

For the positive-polarization ground state of Hubbard--Hund FeO, we obtain
\begin{equation}
S^{hh}_{\mathrm{Fe}|\mathrm O}=0.157890,
\end{equation}
while for Fe$_2$O$_2$,
\begin{equation}
S^{hh}_{\mathrm{Fe}_2|\mathrm O_2}=0.184021.
\end{equation}
In the latter model, separating one Fe center
from the rest of the molecule yields
\begin{equation}
S^{hh}_{\mathrm{Fe}_1|\overline{\mathrm{Fe}_1}}=2.089462,
\end{equation}
whereas separating one O center gives
\begin{equation}
S^{hh}_{\mathrm O_1|\overline{\mathrm O_1}}=0.096778.
\end{equation}
These results illustrate how mode entanglement depends strongly on the spatial partition considered.

{\it Spin-mode entanglement.} Beyond partitioning the modes by which atom they belong to, we can also partition them according to their spin. This allows us to analyze the correlations between orbitals with spin up and those with spin down. Any fermionic pure state with a completely filled spin sector has a unique Fock state in that sector, $|q_{\uparrow}\rangle$, and then the state necessarily factorizes across this partition, implying
\begin{equation}
S_{\uparrow|\downarrow}=0.
\end{equation}
In our models, this applies to the states of Hubbard FeO, as well as to the maximally polarized ground states of Hubbard--Hund FeO and Fe$_2$O$_2$ as per Eqs.~\eqref{eq:Hundfactor} and~\eqref{eq:fe2o2factor}. In intermediate magnetization sectors, where neither spin sector is completely filled, such factorization is no longer guaranteed, and finite spin-mode entanglement may arise.

{\it State dependence within a degenerate ground space.} The examples above concern the choice of mode partition. However, even after fixing the partition, there is still an ambiguity whenever the ground space is degenerate, since the Hamiltonian does not select a unique pure state. Then, since quantities such as the von Neumann entanglement entropy are evaluated on a particular state, different symmetry resolutions of the same ground space can therefore give different values. To illustrate this phenomenon, consider again the Hubbard FeO model $H_{\rm FeO}^{h}$. As mentioned before, every ground state in the local-magnetization basis factorizes between the spin-up and spin-down sectors and therefore satisfies 
\begin{equation}
    S_{\uparrow|\downarrow}=0\,.
\end{equation}
However, the same ground space can also be resolved using total-spin quantum numbers.

For the $S=2$ quintet, a state with magnetization $M$ contains $2-M$ down spins among the
four effective spin-$1/2$ degrees of freedom. The corresponding state is an equal superposition of $\tbinom{4}{2-M}$ spin configurations, which form its Schmidt decomposition across the spin partition. Its entanglement is therefore
\begin{equation}
S_{\uparrow|\downarrow}(M)
=\log_2\binom{4}{2-M},
\end{equation}
implying that for the five members of the quintet one respectively obtains
\begin{equation}
S_{\uparrow|\downarrow}=(0,2,\log_2 6,2,0)
\end{equation}
for $M=2,1,0,-1,-2$. 

Thus, even after the spin partition has been fixed, the entanglement still depends on how the degenerate ground space is resolved. States with definite local magnetizations have zero spin-mode entanglement, whereas states with definite total spin $S$ and magnetization $M$ can exhibit finite entanglement, even though they belong to the same degenerate ground space.

If one instead wants to characterize genuine fermionic correlations beyond the SD description, one has to move beyond bipartite mode entanglement. In the next section, we do this using reduced-density-matrix diagnostics that directly probe departures from a single SD.

\subsection{Fermionic correlations}
\label{sec:fermionic}

To characterize genuine fermionic correlations, or particle entanglement, one can employ the $M$-body reduced density matrices ($M$-RDMs), which probe correlations at different particle orders. Here, for simplicity, we restrict our analysis to the one- and two-body
RDMs. For a system of $n$ modes these are given by~\cite{gigena2015entanglement} 
\begin{subequations}
\begin{eqnarray}
\rho^{(1)}_{ii'} &=& \bra{\Psi} a_{i'}^\dagger a_i\ket{\Psi}\in\mathbb{C}^{n\times n}, \label{eq:1rdm} \\
\rho^{(2)}_{ij,i'j'} &=& \bra{\Psi} a_{i'}^\dagger a_{j'}^\dagger a_j a_i \ket{\Psi}\in\mathbb{C}^{\binom{n}{2}\times \binom{n}{2}}, \label{eq:2rdm}
\end{eqnarray}
\end{subequations}
where $i<j$ and $i'<j'$. They are hermitian, positive semidefinite, and their traces are given by ${\rm Tr}({\bm \rho}^{(1)}) = N$, ${\rm Tr}({\bm \rho}^{(2)}) = \binom{N}{2}$. 

Importantly, unlike the bipartite entropies considered above, the quantities we construct from these RDMs do not require choosing a partition of the modes. Moreover, under a unitary transformation $U\in U(n)$ of the one-particle basis, 
\begin{equation}
\tilde a_i^\dagger=\sum_j U_{ji}a_j^\dagger,
\quad \tilde a_i=\sum_j U_{ji}^*a_j,
\end{equation}
they transform by unitary similarity,
${\bm \rho}^{(1)}\to U^\dagger{\bm \rho}^{(1)}U$ and
${\bm \rho}^{(2)}\to (U\wedge U)^\dagger
{\bm \rho}^{(2)}(U\wedge U)$. Consequently, their eigenvalues are invariant under one-particle basis transformations.

In particular, there always exists a unitary transformation $U$ that diagonalizes $\rho^{(1)}$, such that
\begin{equation}
\tilde{\rho}_{ij}^{(1)} = \langle\Psi|\tilde a_i^\dagger\tilde a_j|\Psi\rangle
=n_i\delta_{ij},
\qquad \sum_i n_i=N.
\end{equation}
The eigenvalues $n_i\in[0,1]$, known as the
natural occupations, give the average occupation
of each natural orbital. 

For pure states, the state is a single SD if and only if all natural occupations are either $0$ or $1$. In this case, one finds
\begin{equation}
    (\rho^{(1)})^2=\rho^{(1)}\,,\label{eq:idempot}
\end{equation}
Thus, we can quantify the particle entanglement through the one-body von Neumann entropy
\begin{equation}
S^{(1)}=-\Tr\left({\bm \rho}^{(1)}
\log_2{\bm \rho}^{(1)}\right)
\label{eq:E1}
\end{equation}
as before, which vanishes if and only if the pure state is a single SD. 

Moreover, departures from Eq.~\eqref{eq:idempot} directly probe fermionic correlations beyond the SD manifold and can be quantified through the idempotency defect
\begin{equation}
\delta^{(1)}=\|{(\bm \rho}^{(1)})^2-{\bm \rho}^{(1)}\|_F.
\label{eq:idempotency}
\end{equation}
We also consider the following quantities constructed from the same natural occupation numbers~\cite{tarabunga2026computable,diaz2023showcasing}:
\begin{align}
M^{[1]}=\sum_i h_2(n_i),\quad 
M^{[2]}=4\sum_i n_i(1-n_i),
\end{align}
where
\begin{equation}
h_2(x)=-x\log_2x-(1-x)\log_2(1-x).
\end{equation}
All of these previous quantities vanish when the natural occupations are restricted to zero and one, and are therefore zero for a pure fixed-particle-number SD.

Then, regarding the two-body RDM, we know that for a SD, this matrix is completely determined by the one-body RDM through Wick's theorem. Specifically, we have
\begin{equation}
\rho_{ij,i'j'}^{(2)}=\rho_{ii'}^{(1)}\rho_{jj'}^{(1)}-\rho_{ij'}^{(1)}\rho_{ji'}^{(1)}.
\end{equation}
We can therefore isolate correlations that cannot be reconstructed from the one-body data by defining the two-body cumulant matrix
\begin{equation}
\kappa_{ij,i'j'}^{(2)}=\rho_{ij,i'j'}^{(2)}-(\rho_{ii'}^{(1)}\rho_{jj'}^{(1)}-\rho_{ij'}^{(1)}\rho_{ji'}^{(1)})
\label{eq:cumulant}
\end{equation}
For a SD, ${\bm \kappa}={\bf 0}$, and then, a finite cumulant therefore provides a direct two-body signature of the failure of Wick factorization.

\subsubsection{The FeO models}

Due to the symmetries of both the Hubbard and Hubbard--Hund FeO Hamiltonians, their eigenstates can be chosen with definite particle numbers $N_{\pi_y\pi_z}$ and local magnetizations $M_{\pi_y\pi_z}$. For such states, the one-body RDM is block diagonal with respect to the $4$ orbital groups~\cite{cianciulli2024bipartite},
\begin{equation}
{\bm\rho}^{(1)}
=\bigoplus_{\pi_y,\pi_z=\pm}
{\bm\rho}^{(1)}_{\pi_y\pi_z},
\label{orbdesc}
\end{equation}
with block dimensions $(6,4,4,2)$ and traces $(5,3,3,1)$. Then, each block can be further decomposed into spin sectors as
\begin{equation}
{\bm\rho}^{(1)}_{\pi_y\pi_z}
={\bm\rho}^{(1)}_{\pi_y\pi_z\uparrow}
\oplus{\bm\rho}^{(1)}_{\pi_y\pi_z\downarrow},
\label{spindesc}
\end{equation}
with block dimensions $|\mathcal C_{\pi_y\pi_z}|$
and traces $N_{\pi_y\pi_z\uparrow}$,
$N_{\pi_y\pi_z\downarrow}$, respectively ( $N_{\pi_y\pi_z\uparrow(\downarrow)}
=N_{\pi_y\pi_z}/2\pm M_{\pi_y\pi_z}$).

In both models, all ground states belong to the same particle-number sector of  Eq.~\eqref{eq:GSsectoroccupations}, so the first decomposition in~\eqref{orbdesc} holds throughout the ground space. We now examine the fermionic correlations of the ground states in each approximation separately.

{\it Hubbard FeO}. On the one hand, we note that when resolving the ground space by local magnetizations, all $16$ ground states are exact SD. Consequently,
for all these states, the one-body RDM has $12$ unit eigenvalues and $4$ zero eigenvalues,
\begin{equation}
\spec\bm\rho^{(1)}
=\{1^{\times12},0^{\times4}\}\,.
\end{equation}
It follows that
\begin{equation}
S_{\rm FeO,h}^{(1)}=\delta^{(1)}=M^{[1]}=M^{[2]}=0, \qquad \bm\kappa_{\rm FeO,h}^{(2)}=\bm 0.
\end{equation}
For instance, for the positive-polarization ground state, the spin-up blocks are completely filled, implying
\begin{equation}
\bm\rho_{\pi_y\pi_z,\uparrow}^{(1)}
=\mathbb I_{|\mathcal C_{\pi_y\pi_z}|},
\label{rhoful}
\end{equation}
while the spin-down blocks have spectra
\begin{equation}
\begin{aligned}
\spec\bm\rho_{++\downarrow}^{(1)}
&=\{1^{\times2},0\},\\
\spec\bm\rho_{\pm\mp\downarrow}^{(1)}
&=\{1,0\},\\
\spec\bm\rho_{--\downarrow}^{(1)}
&=\{0\}.
\end{aligned}
\end{equation}

On the other hand, the coupled-spin states with fixed total magnetization within the degenerate ground space can exhibit nontrivial fermionic correlations. In every state of the ground manifold, $4$ spatial orbitals are doubly occupied and $4$ are singly occupied. The doubly occupied orbitals contribute eight unit natural occupations, while the remaining four electrons occupy one active spatial orbital in each group. Thus, for an arbitrary state in the ground manifold, the $4$ one-body RDM blocks have spectra
\begin{equation}
\begin{aligned}
\spec\bm\rho_{\scriptscriptstyle ++}^{(1)}
&=\{1^{\times4},p_{\scriptscriptstyle ++},1-p_{\scriptscriptstyle ++}\},\\
\spec\bm\rho_{\scriptscriptstyle \pm \mp}^{(1)}
&=\{1^{\times2},p_{\scriptscriptstyle \pm \mp},1-p_{\scriptscriptstyle \pm \mp}\},\\
\spec\bm\rho_{\scriptscriptstyle --}^{(1)}
&=\{p_{\scriptscriptstyle --},1-p_{\scriptscriptstyle --}\},
\end{aligned}
\end{equation}
where $0\leq p_{\pi_y\pi_z}\leq1$ are the
eigenvalues of the reduced spin states of
the four active orbitals, with the complementary
eigenvalues given by $1-p_{\pi_y\pi_z}$.

For instance, for the $S=2$ quintet, the states are fully symmetric under permutations of the $4$ effective spins, implying 
\begin{equation}
    p_{\pi_y\pi_z}=(2+M)/4
\end{equation}
for all the $4$ orbital groups. Then, we find 
\begin{equation}
S_{\rm FeO,h}^{(1)}(S=2,M)
=-\sum_{\sigma=\pm1}(2+\sigma M)
\log_2\frac{2+\sigma M}{4}.
\end{equation}
Hence, $S_{\rm FeO,h}^{(1)}(S=2,M=\pm2)=0$, corresponding to Slater determinants, while
$S_{\rm FeO,h}^{(1)}(S=2,M=0)=4$. Moreover, we also obtain
\begin{align}
M^{[1]}(S=2,M)&=8h_2\left(\frac{2+ M}{4}\right),\\
M^{[2]}(S=2,M)&=8-2M^2,\\
\delta^{(1)}(S=2,M)&=\frac{4-M^2}{4\sqrt2}.
\label{eq:quintMeasures}
\end{align}
Here we can also see that the extremal states with $M=\pm2$ are Slater determinants. By contrast, the $M=0$ state has
\begin{equation}
M^{[1]}=8,
\quad
M^{[2]}=8,\quad \delta^{(1)}=\frac{1}{\sqrt2}.
\end{equation}

{\it Hubbard-Hund FeO}. We now consider the positive-polarization ground state of the Hubbard--Hund model. Since all $8$ spin-up
orbitals are occupied, we obtain Eq.~\eqref{rhoful} for the spin-up block, while the spectra of the spin-down one-body RDMs take the form
\begin{equation}
\begin{aligned}
\spec\bm\rho_{\scriptscriptstyle ++\downarrow}^{(1)}
&=\{1-\alpha_{\scriptscriptstyle ++},1-\beta_{\scriptscriptstyle ++},\alpha_{\scriptscriptstyle ++}+\beta_{\scriptscriptstyle ++}\},\\
\spec\bm\rho_{\scriptscriptstyle \pm \mp \downarrow}^{(1)}
&=\{1-\alpha_{\scriptscriptstyle \pm \mp},\alpha_{\scriptscriptstyle \pm \mp}\},\\
\spec\bm\rho_{\scriptscriptstyle --\downarrow}^{(1)}
&=\{0\}.
\end{aligned}
\end{equation}
Thus, only $4$ independent natural occupations
are required to characterize the entire one-body RDM spectrum. We obtain 
$\alpha_{++}=6.22919\times10^{-10}$, $\beta_{++}=9.67\times10^{-17}$,
$\alpha_{+-}=1.91384061\times10^{-7}$,
and $\alpha_{-+} = 1.91342574\times10^{-7}$. Hence, these values are extremely close to zero or one, yielding the one-body entanglement measures
\begin{align}
S^{(1)}&=9.11\!\times\!10^{-6},\quad
M^{[1]}=1.82\!\times\!10^{-5},\nonumber\\
M^{[2]}&=3.06\!\times\!10^{-6},\quad
\delta^{(1)}=3.83\!\times\!10^{-7},\nonumber\\
&\|\bm\kappa_{\rm FeO,hh}^{(2)}\|_F\nonumber
=6.19\times10^{-4}.
\end{align}
Nevertheless, the state is not an exact Slater determinant, and all nontrivial fermionic
correlations are confined to the minority-spin sector.  

The negative-polarization ground state exhibits analogous properties, with spin-up and spin-down interchanged. As in the Hubbard model, arbitrary superpositions of the two degenerate ground states need not have definite local magnetizations. However, the
spin-block decomposition \eqref{spindesc} of the one-body RDM remains valid, since their magnetizations differ by more than a one-body operator can connect.

\subsubsection{The Fe$_2$O$_2$ model}

Due to the symmetries of the Hubbard--Hund Fe$_2$O$_2$ Hamiltonian, its eigenstates can
be chosen with definite particle numbers $N_\pm$, local magnetizations $M_\pm$, and
occupation parities $P_{\pi_x\pi_y\pi_z}$. As before, for such states, the one-body RDM is block diagonal with respect to the eight orbital groups,
\begin{equation}
\bm\rho^{(1)}
=\bigoplus_{\pi_x,\pi_y,\pi_z=\pm}
\bm\rho_{\pi_x\pi_y\pi_z}^{(1)},
\label{rhopides}
\end{equation}
and each block further decomposes into spin sectors,
\begin{equation}
\bm\rho_{\pi_x\pi_y\pi_z}^{(1)}
=\bm\rho_{\pi_x\pi_y\pi_z\uparrow}^{(1)}
\oplus
\bm\rho_{\pi_x\pi_y\pi_z\downarrow}^{(1)}.
\label{rhospinpides}
\end{equation}
The spin-resolved blocks have dimensions $|\mathcal C_{\pi_x\pi_y\pi_z}|$, while their traces satisfy $\Tr(\bm\rho_{\pi_x\pi_y\pi_z\uparrow (\downarrow)}^{(1)}) =\langle\hat N_{\pi_x\pi_y\pi_z\uparrow (\downarrow)}\rangle$. Unlike in FeO, these traces are not generally fixed by the symmetries, since the individual particle numbers are not conserved. 

We now consider the positive-polarization ground state. As shown above, the effective Hamiltonian in this sector recovers the conservation of the individual particle numbers
$N_{\pi_x\pi_y\pi_z}$. Consequently, the decompositions \eqref{rhopides} and
\eqref{rhospinpides} hold with fixed block traces, determined by the occupation numbers
in Eq.~\eqref{Nsector}. Indeed, since all $16$ spin-up orbitals are occupied, we have
\begin{equation}
\bm\rho_{\pi_x\pi_y\pi_z\uparrow}^{(1)}
=\mathbb I_{|\mathcal C_{\pi_x\pi_y\pi_z}|},
\end{equation}
while the spin-down blocks have fixed traces
\begin{equation}
    \Tr(\bm\rho_{\pi_x\pi_y\pi_z\downarrow}^{(1)})
= N_{\pi_x\pi_y\pi_z} -|\mathcal C_{\pi_x\pi_y\pi_z}|\,.
\end{equation}
Hence, the spectra of the spin-down one-body RDM blocks take the following forms,
\begin{subequations}
\begin{eqnarray}
\operatorname{spec}\!\rho^{(1)}_{\scriptstyle ++-\downarrow}
&\!\!=\!\!&\{1,1\},\\
\operatorname{spec}\!\rho^{(1)}_{\scriptstyle +++\downarrow}
&\!\!=\!\!&\{1 \!-\! \alpha_{\scriptstyle +++},\!1 \!-\! \beta_{\scriptstyle +++},
\!\alpha_{\scriptstyle +++} \!+\! \beta_{\scriptstyle +++}\}, \;\;\;\;\;\;\; \\
\operatorname{spec}\!\rho^{(1)}_{\scriptstyle -++\downarrow}
&\!\!=\!\!&\{1-\alpha_{\scriptstyle -++}-\beta_{\scriptstyle -++},
\!\alpha_{\scriptstyle -++},\!\beta_{\scriptstyle -++}\},\\
\operatorname{spec}\!\rho^{(1)}_{\scriptstyle +-\pm\downarrow}
&\!\!=\!\!&\{1-\alpha_{\scriptstyle +-\pm},\!\alpha_{\scriptstyle +-\pm}\},\\
\operatorname{spec}\!\rho^{(1)}_{\scriptstyle --+\downarrow}
&\!\!=\!\!&\{1-\alpha_{\scriptstyle --+},\!\alpha_{\scriptstyle --+}\},\\
\operatorname{spec}\!\rho^{(1)}_{\scriptstyle -\pm-\downarrow}
&\!\!=\!\!&\{0\}.
\end{eqnarray}
\end{subequations}
Thus, only seven independent natural occupations are required to characterize the entire one-body RDM spectrum. In particular, the two orbitals in $\mathcal C_{++-}$ are fully occupied, while those in $\mathcal C_{-+-}$ and $\mathcal C_{---}$ are exactly empty, consistently with the ground-state structure previously established. Numerically, we obtain
$\alpha_{+++}=2.14448\times10^{-9}$,
$\beta_{+++}=1.47\times10^{-15}$,
$\alpha_{-++}=3.64641\times10^{-8}$,
$\beta_{-++}=2.51\times10^{-19}$,
$\alpha_{+-+}=5.25097\times10^{-8}$,
$\alpha_{+--}=3.82848\times10^{-8}$,
and $\alpha_{--+}=5.34495\times10^{-8}$, yielding
\begin{align}
S_{\rm Fe_2O_2,hh}^{(1)}&\!=\!4.73\!\times\!10^{-6},\quad
M_{\rm Fe_2O_2,hh}^{(1)[1]}\!=\!9.46\!\times\!10^{-6},\nonumber\\
M_{\rm Fe_2O_2,hh}^{(1)[2]}&\!=\!1.46\!\times\!10^{-6},\quad \delta_{\rm Fe_2O_2,hh}^{(1)}\!=\!1.30\!\times\!10^{-7},\nonumber\\
&\|\bm\kappa_{\rm Fe_2O_2,hh}^{(2)}\|_F\nonumber
=4.27\times10^{-4}.
\end{align}
Thus, despite its larger orbital space and its substantially larger entanglement across some spatial partitions, Fe$_2$O$_2$ is slightly closer to the Slater-determinant manifold than the Hubbard-Hund FeO ground state. The two examples therefore show that increasing spatial entanglement does not necessarily imply an increase in fermionic non-Gaussianity.

Finally, we note that the negative-polarization ground state has analogous properties, with spin-up and spin-down interchanged. As before, superpositions of the two degenerate ground states need not have definite magnetizations, but the spin-block decompositions in Eq.~\eqref{rhopides} and~\eqref{rhospinpides} of the one-body RDM remains valid.

\section{Fermionic transformation classes and conserved operators}
\label{sec:CO}

Let us slightly change gears. So far, we have studied static properties of a chosen ground state. We asked, for instance, how strongly the fermionic reduced density matrices depart from those of a SD. In this sense, the questions considered so far are analogous to asking: \textit{How much entanglement is present in a fixed multipartite state?} One can instead ask a more dynamical question in the language of quantum resource theories~\cite{chitambar2019quantum,brandao2015reversible}. Namely, \textit{given a state, which other states can be reached from it using a prescribed set of free transformations?}

More concretely, and for ease of comparison, consider the standard resource theory of multipartite entanglement. There, local operations and classical communication (LOCC) define the free transformations~\cite{chitambar2014everything}, while considering a single invertible stochastic branch leads to the familiar stochastic LOCC (SLOCC) action~\cite{dur2000three,miyake2003classification}. The corresponding question is then \textit{given a pure state, what other states can I reach via SLOCC?} Answering this question naturally leads to SLOCC orbits and transformation classes. States in the same orbit can be interconverted by invertible SLOCC transformations, while states in different orbits belong to different SLOCC classes~\cite{verstraete2002four,dietrich2022classification,jia2015classification}.

Thus, we can identify two different tasks, whereby entanglement measures ask how much entanglement a state possesses, while the resource-theoretic viewpoint asks which states can be transformed into one another using the allowed free operations. We apply the same logic here to fixed-particle-number fermionic states. The natural free unitary transformations are passive fermionic linear-optical transformations, which act as unitary rotations of the one-particle modes. Allowing invertible stochastic branches leads to their complexification from $U(n)$ to $GL(n,\mathbb C)$~\cite{diaz2025unified}. Thus, for an $N$-fermion state in $n$ modes, we consider transformations of the form
\begin{equation}
\ket{\psi}\longmapsto\left(\bigwedge^N g\right)\ket{\psi},
\qquad
g\in GL(n,\mathbb C)\,.
\label{eq:GLaction}
\end{equation}
The set of states obtained by varying $g$ defines the $GL(n,\mathbb C)$ orbit of $\ket{\psi}$. States belonging to the same orbit are interconvertible under invertible branches associated with particle-number-preserving one-body transformations and therefore belong to the same fermionic transformation class.

Classifying and reconstructing the full orbit of a state can be a daunting task. However, one can find necessary conditions for one state to be interconvertible with another using the state-dependent conserved-operator formalism recently introduced in Ref.~\cite{petrovich2026symmetry}. Here, we identify the one-body operators that leave a state invariant up to an overall scalar and use the resulting algebra as a fingerprint of its transformation class. States belonging to the same $GL(n,\mathbb C)$ orbit must have conserved-operator algebras related by the corresponding change of one-particle basis, and in particular must have the same stabilizer dimension.

Specifically, for $u\in\mathfrak{gl}(n,\mathbb C)$, let
\begin{equation}
Q_u=\sum_{i,j}u_{ij}a_i^\dagger a_j
\label{eq:OX}
\end{equation}
be the corresponding infinitesimal one-body transformation. Acting with $Q_u$ on $\ket{\psi}$ gives the tangent vector generated by moving along the $GL(n,\mathbb C)$ orbit through $\ket{\psi}$. The directions for which this tangent vector is proportional to $\ket{\psi}$ itself do not move the corresponding projective state and therefore define the infinitesimal stabilizer of $[\psi]$. From this geometrical picture, we call $Q$ a conserved operator of $\ket{\psi}$ when
\begin{equation}
Q_u\ket{\psi}=\chi_\psi(Q_u)\ket{\psi},
\label{eq:COcondition}
\end{equation}
where $\chi_\psi(Q_u)$ is the corresponding eigenvalue.

The coefficient matrices satisfying Eq.~\eqref{eq:COcondition} form the projective stabilizer algebra
\begin{equation}
\pstab{[\psi]}=
\left\{
u\in\mathfrak{gl}(n,\mathbb C)
\ \middle|\
Q_u\ket{\psi}=\chi_\psi(Q_u)\ket{\psi}
\right\}.
\label{eq:pstab}
\end{equation}

The connection with interconvertibility follows directly from how this algebra transforms under $GL(n,\mathbb C)$. Suppose that two states belong to the same orbit, such that $
\ket{\phi}\propto\left(\bigwedge^N g\right)\ket{\psi}$ for some $g\in GL(n,\mathbb C)$. The corresponding infinitesimal generators transform as $
Q_{gug^{-1}}=
\left(\bigwedge^N g\right)
Q_u \left(\bigwedge^N g\right)^{-1}$. Therefore, if $Q\in\pstab{[\psi]}$, then $gQg^{-1}\in\pstab{[\phi]}$, and hence
\begin{equation}
\pstab{[\phi]} =
g\,\pstab{[\psi]}\,g^{-1}.
\label{eq:pstabconjugation}
\end{equation}
States in the same $GL(n,\mathbb C)$ orbit must therefore have conjugate, and in particular isomorphic, conserved-operator algebras. In particular, they must also have the same number of linearly independent conserved operators,
\begin{equation}
\dim_{\mathbb C}\pstab{[\phi]}
= \dim_{\mathbb C}\pstab{[\psi]}.
\end{equation}
Thus, a difference in stabilizer dimension immediately rules out interconvertibility, while nonisomorphic stabilizer algebras can distinguish states even when their dimensions agree. 

Moreover, defining $G=GL(n,\mathbb C)$ and denoting by $G\cdot{[\psi]}$ the orbit of $\ket{\psi}$, one finds
\begin{equation}
\dim_{\mathbb C}G \cdot{[\psi]}=n^2-\dim_{\mathbb C}\pstab{[\psi]}.
\label{eq:orbitdim}
\end{equation}
Thus, the more conserved operators a state possesses, the fewer independent directions are available to move it along its orbit, and consequently the smaller the orbit becomes. Importantly, among fixed-particle-number fermionic pure states, SDs maximize $\dim_{\mathbb C}\pstab_{[\psi]}$. For an $N$-particle SD, $|\psi_{\rm SD}\rangle$, in $n$ modes,
\begin{subequations}
\begin{eqnarray}
\dim_{\mathbb C} \pstab[\psi _{\rm SD}] &=& n^2-N(n-N), \label{eq:SDpdim}  \\ 
\dim_{\mathbb C} G \cdot [\psi_{\rm SD}] &=& N(n-N). \label{eq:SDOdim}
\end{eqnarray}
\end{subequations}
In this sense, SDs have the smallest transformation classes under invertible one-particle operations.

Importantly, conserved operators provide a particularly appealing framework to classify state interconvertibility, as they can be obtained from the same one- and two-body RDMs used in the previous section. Defining
\begin{equation}
O_{ij}=a_i^\dagger a_j,
\qquad
\Delta O_{ij}=O_{ij}-\langle O_{ij}\rangle,
\end{equation}
we construct the covariance matrix
\begin{equation}
C_{ij,kl}=
\langle
\Delta O_{ij}^\dagger
\Delta O_{kl}
\rangle.
\label{eq:covCO}
\end{equation}
Using the fermionic anticommutation relations, every entry of $\bf C$ can be expressed in terms of $\bm \rho^{(1)}$ and $\bm \rho^{(2)}$. Its null space gives the coefficient matrices satisfying Eq.~\eqref{eq:COcondition} \cite{petrovich2026symmetry}.

Taken together, conserved operators give us a different and complementary way of comparing the ground states introduced above. The measures in Sec.~\ref{sec:fermionic} tell us how strongly a state departs from a Slater determinant, while the conserved-operator algebra tells us which infinitesimal one-particle transformations leave the state unchanged and, through Eq.~\eqref{eq:orbitdim}, the dimension of its $GL(n,\mathbb C)$ orbit. As we now show, these two notions can behave very differently.

\subsection{Hubbard FeO}

We begin with the local-magnetization ground states of Hubbard FeO, which are exact Slater determinants. For the present case, $n=16$ and $N=12$, and Eqs.~\eqref{eq:SDpdim} and~\eqref{eq:SDOdim} give
\begin{equation}
\dim_{\mathbb C} \pstab[\psi]=208,
\qquad
\dim_{\mathbb C} G\cdot [\psi]=48.
\label{eq:HubbardSDorbit}
\end{equation}
Every one of the sixteen local-magnetization ground states has these values.

The coupled-spin states of the same ground space have a very different conserved-operator structure. For the $S=2$ quintet, see Table~\ref{tab:quintCO}. 
\begin{table}[h!]
\caption{Conserved-operator data for the $S=2$ quintet contained in the Hubbard FeO ground space.}
\label{tab:quintCO}
\centering
\begin{tabular}{c c c}
\toprule
$M$ & $\dim_{\mathbb C} \pstab[\psi]$ & $\dim_{\mathbb C} G\cdot [\psi]$\\
\midrule
$\pm2$ & 208 & 48\\
$\pm1$ & 160 & 96\\
$0$ & 145 & 111\\
\bottomrule
\end{tabular}
\end{table}

As expected, the extremal states with $M=\pm2$ match Eq.~\eqref{eq:HubbardSDorbit} as they  are the Slater determinants. Moving toward the center of the quintet substantially changes the transformation class. At $M=0$, the stabilizer dimension decreases from $208$ to $145$, while the orbit dimension increases from $48$ to $111$. Thus, the same exact ground space contains states lying on different $GL(16,\mathbb C)$ orbits.

This provides another manifestation of the ambiguity associated with resolving a degenerate ground space. The Hamiltonian and the energy are unchanged, yet different symmetry resolutions select states with different transformation classes under invertible one-particle operations.

\subsection{Hubbard-Hund FeO}

The ground states of the Hubbard-Hund FeO provide a different comparison. In Sec.~\ref{sec:fermionic}, we found that its natural occupations lie extremely close to zero and one, with $
\delta^{(1)}=3.82728\times10^{-7}$. According to the one-body RDM, the state is therefore extremely close to the Slater-determinant manifold. Its conserved-operator structure tells a rather different story.

The covariance matrix in Eq.~\eqref{eq:covCO} gives
\begin{equation}
\dim_{\mathbb C} \pstab{[\Psi_+^{\rm FeO}]}=152,
\qquad
\dim_{\mathbb C} G \cdot [\Psi_+^{\rm FeO}]=104.
\label{eq:Hundorbit}
\end{equation}
Thus, although the natural occupations differ from zero and one only at the level of approximately $10^{-7}$, the stabilizer and orbit dimensions are far from the Slater values \eqref{eq:HubbardSDorbit}. Numerical proximity to the Slater-determinant manifold therefore does not imply proximity of the corresponding transformation class.

\subsection{Fe$_2$O$_2$}

The distinction made in the previous section is also present for Fe$_2$O$_2$. Its one-body idempotency defect is $\delta^{(1)}=1.29723\times10^{-7}$, which is even smaller than the value found for Hubbard-Hund FeO. Nevertheless, the minority-spin state has a conserved-operator structure that differs strongly from that of a Slater determinant. Indeed, for the eight-electron minority-spin state in sixteen modes, the covariance matrix gives a numerically well-resolved stabilizer dimension
\begin{equation}
\dim_{\mathbb C}\pstab [\Psi_+^{\rm Fe_2O_2}]=610,\quad \dim_{\mathbb C}  G \cdot [\Psi_+^{\rm Fe_2O_2}] =414.
\label{eq:Fe2pdown}
\end{equation}
A $24$-particle SD in the same $32$-mode space would instead have
\begin{equation}
\dim_{\mathbb C}\pstab [\psi_{\rm SD}]=832,
\qquad
\dim_{\mathbb C} G \cdot [\psi_{\rm SD}]=192.
\end{equation}
Thus, despite an idempotency defect of only order $10^{-7}$, the numerically resolved orbit has dimension $414$, compared with the Slater-determinant value $192$.

\section{Discussion}
\label{sec:discussion}

In this work we have explored the rich structure that can emerge within degenerate ground-state manifolds with nontrivial symmetry structure through the optics of mode and particle entanglement. Indeed, the iron-oxide models studied here show that these notions can capture rather different features of the same state. Large entanglement across a physically motivated mode partition need not imply strong correlations beyond the SD description, while states whose natural occupations are extremely close to those of a SD can nevertheless belong to very different fermionic orbits. This makes it useful to keep the different notions of correlation separate rather than trying to compress them into a single measure.

In the broader context of our companion work on the quantum simulation of these iron-oxide models~\cite{staros2026towards}, these results also give some guidance on how one should think about simple representations of the ground state. This is particularly relevant for dynamical simulations, where the ground state is often only the starting point of the calculation and must first be prepared before the dynamics can be applied. In a degenerate manifold, different symmetry resolutions can select states with substantially different structure, and some of these may admit much simpler representations than others. The choice of symmetry sector and state-preparation prescription can therefore become part of the algorithmic design.

Moving forward, it will be important to connect these structural diagnostics more directly to concrete simulation resources. In particular, one would like to understand when symmetry resolution, near-Slater structure, or conserved-operator information can be converted into reductions in state-preparation depth, sampling cost, or the resources required for subsequent dynamical simulations of correlated iron-oxide systems. Crucially, our work illustrated the well-known fact that careful symmetry analysis allows one to reduce the search space for, e.g., ground states, significantly. Indeed, already for Fe$_2$O$_2$ we were able to narrow down the dimension of the relevant Hilbert space by three orders of magnitude, and that of the ground state block by more than $5$ orders of magnitude. 

\section{Acknowledgments}
We thank Rahul Hingorani for useful and productive discussions. This work was supported by the Laboratory Directed Research and Development (LDRD) program of Los Alamos National Laboratory (LANL) under project number 20260043DR.

\section{Artificial Intelligence Disclosure}
The idea for this project was created by humans. ChatGPT 5.6 Sol and Astra were used for the explicit symmetry analysis and for polishing the main text. All results were verified by the authors.

\bibliography{quantum}

\begin{thebibliography}{46}%
\makeatletter
\providecommand \@ifxundefined [1]{%
 \@ifx{#1\undefined}
}%
\providecommand \@ifnum [1]{%
 \ifnum #1\expandafter \@firstoftwo
 \else \expandafter \@secondoftwo
 \fi
}%
\providecommand \@ifx [1]{%
 \ifx #1\expandafter \@firstoftwo
 \else \expandafter \@secondoftwo
 \fi
}%
\providecommand \natexlab [1]{#1}%
\providecommand \enquote  [1]{``#1''}%
\providecommand \bibnamefont  [1]{#1}%
\providecommand \bibfnamefont [1]{#1}%
\providecommand \citenamefont [1]{#1}%
\providecommand \href@noop [0]{\@secondoftwo}%
\providecommand \href [0]{\begingroup \@sanitize@url \@href}%
\providecommand \@href[1]{\@@startlink{#1}\@@href}%
\providecommand \@@href[1]{\endgroup#1\@@endlink}%
\providecommand \@sanitize@url [0]{\catcode `\\12\catcode `\$12\catcode `\&12\catcode `\#12\catcode `\^12\catcode `\_12\catcode `\%12\relax}%
\providecommand \@@startlink[1]{}%
\providecommand \@@endlink[0]{}%
\providecommand \url  [0]{\begingroup\@sanitize@url \@url }%
\providecommand \@url [1]{\endgroup\@href {#1}{\urlprefix }}%
\providecommand \urlprefix  [0]{URL }%
\providecommand \Eprint [0]{\href }%
\providecommand \doibase [0]{https://doi.org/}%
\providecommand \selectlanguage [0]{\@gobble}%
\providecommand \bibinfo  [0]{\@secondoftwo}%
\providecommand \bibfield  [0]{\@secondoftwo}%
\providecommand \translation [1]{[#1]}%
\providecommand \BibitemOpen [0]{}%
\providecommand \bibitemStop [0]{}%
\providecommand \bibitemNoStop [0]{.\EOS\space}%
\providecommand \EOS [0]{\spacefactor3000\relax}%
\providecommand \BibitemShut  [1]{\csname bibitem#1\endcsname}%
\let\auto@bib@innerbib\@empty
\bibitem [{\citenamefont {Amico}\ \emph {et~al.}(2008)\citenamefont {Amico}, \citenamefont {Fazio}, \citenamefont {Osterloh},\ and\ \citenamefont {Vedral}}]{amico2008entanglement}%
  \BibitemOpen
  \bibfield  {author} {\bibinfo {author} {\bibfnamefont {L.}~\bibnamefont {Amico}}, \bibinfo {author} {\bibfnamefont {R.}~\bibnamefont {Fazio}}, \bibinfo {author} {\bibfnamefont {A.}~\bibnamefont {Osterloh}},\ and\ \bibinfo {author} {\bibfnamefont {V.}~\bibnamefont {Vedral}},\ }\bibfield  {title} {\bibinfo {title} {Entanglement in many-body systems},\ }\href {https://doi.org/10.1103/RevModPhys.80.517} {\bibfield  {journal} {\bibinfo  {journal} {Reviews of modern physics}\ }\textbf {\bibinfo {volume} {80}},\ \bibinfo {pages} {517} (\bibinfo {year} {2008})}\BibitemShut {NoStop}%
\bibitem [{\citenamefont {Eisert}\ \emph {et~al.}(2010)\citenamefont {Eisert}, \citenamefont {Cramer},\ and\ \citenamefont {Plenio}}]{eisert2010colloquium}%
  \BibitemOpen
  \bibfield  {author} {\bibinfo {author} {\bibfnamefont {J.}~\bibnamefont {Eisert}}, \bibinfo {author} {\bibfnamefont {M.}~\bibnamefont {Cramer}},\ and\ \bibinfo {author} {\bibfnamefont {M.~B.}\ \bibnamefont {Plenio}},\ }\bibfield  {title} {\bibinfo {title} {Colloquium: Area laws for the entanglement entropy},\ }\href {https://doi.org/10.1103/RevModPhys.82.277} {\bibfield  {journal} {\bibinfo  {journal} {Reviews of modern physics}\ }\textbf {\bibinfo {volume} {82}},\ \bibinfo {pages} {277} (\bibinfo {year} {2010})}\BibitemShut {NoStop}%
\bibitem [{\citenamefont {De~Chiara}\ and\ \citenamefont {Sanpera}(2018)}]{de2018genuine}%
  \BibitemOpen
  \bibfield  {author} {\bibinfo {author} {\bibfnamefont {G.}~\bibnamefont {De~Chiara}}\ and\ \bibinfo {author} {\bibfnamefont {A.}~\bibnamefont {Sanpera}},\ }\bibfield  {title} {\bibinfo {title} {Genuine quantum correlations in quantum many-body systems: a review of recent progress},\ }\href {https://doi.org/10.1088/1361-6633/aabf61} {\bibfield  {journal} {\bibinfo  {journal} {Reports on Progress in Physics}\ }\textbf {\bibinfo {volume} {81}},\ \bibinfo {pages} {074002} (\bibinfo {year} {2018})}\BibitemShut {NoStop}%
\bibitem [{\citenamefont {Laflorencie}(2016)}]{laflorencie2016quantum}%
  \BibitemOpen
  \bibfield  {author} {\bibinfo {author} {\bibfnamefont {N.}~\bibnamefont {Laflorencie}},\ }\bibfield  {title} {\bibinfo {title} {Quantum entanglement in condensed matter systems},\ }\href {https://doi.org/10.1016/j.physrep.2016.06.008} {\bibfield  {journal} {\bibinfo  {journal} {Physics Reports}\ }\textbf {\bibinfo {volume} {646}},\ \bibinfo {pages} {1} (\bibinfo {year} {2016})}\BibitemShut {NoStop}%
\bibitem [{\citenamefont {Tarabunga}(2024)}]{tarabunga2024critical}%
  \BibitemOpen
  \bibfield  {author} {\bibinfo {author} {\bibfnamefont {P.~S.}\ \bibnamefont {Tarabunga}},\ }\bibfield  {title} {\bibinfo {title} {Critical behaviors of non-stabilizerness in quantum spin chains},\ }\href {https://doi.org/10.22331/q-2024-07-17-1413} {\bibfield  {journal} {\bibinfo  {journal} {Quantum}\ }\textbf {\bibinfo {volume} {8}},\ \bibinfo {pages} {1413} (\bibinfo {year} {2024})}\BibitemShut {NoStop}%
\bibitem [{\citenamefont {Sarkis}\ and\ \citenamefont {Tkatchenko}(2025)}]{sarkis2025molecules}%
  \BibitemOpen
  \bibfield  {author} {\bibinfo {author} {\bibfnamefont {M.}~\bibnamefont {Sarkis}}\ and\ \bibinfo {author} {\bibfnamefont {A.}~\bibnamefont {Tkatchenko}},\ }\bibfield  {title} {\bibinfo {title} {Are molecules magical? non-stabilizerness in molecular bonding},\ }\bibfield  {journal} {\bibinfo  {journal} {arXiv preprint arXiv:2504.06673}\ }\href {https://doi.org/10.48550/arXiv.2504.06673} {10.48550/arXiv.2504.06673} (\bibinfo {year} {2025})\BibitemShut {NoStop}%
\bibitem [{\citenamefont {Sierant}\ \emph {et~al.}(2025)\citenamefont {Sierant}, \citenamefont {Stornati},\ and\ \citenamefont {Turkeshi}}]{deside2025detecting}%
  \BibitemOpen
  \bibfield  {author} {\bibinfo {author} {\bibfnamefont {P.}~\bibnamefont {Sierant}}, \bibinfo {author} {\bibfnamefont {P.}~\bibnamefont {Stornati}},\ and\ \bibinfo {author} {\bibfnamefont {X.}~\bibnamefont {Turkeshi}},\ }\bibfield  {title} {\bibinfo {title} {Fermionic magic resources of quantum many-body systems},\ }\bibfield  {journal} {\bibinfo  {journal} {arXiv preprint arXiv:2506.00116}\ }\href {https://doi.org/10.48550/arXiv.2506.00116} {10.48550/arXiv.2506.00116} (\bibinfo {year} {2025})\BibitemShut {NoStop}%
\bibitem [{\citenamefont {Moca}\ \emph {et~al.}(2025)\citenamefont {Moca}, \citenamefont {Sticlet}, \citenamefont {D{\'o}ra}, \citenamefont {Valli}, \citenamefont {Szombathy},\ and\ \citenamefont {Zar{\'a}nd}}]{moca2025non}%
  \BibitemOpen
  \bibfield  {author} {\bibinfo {author} {\bibfnamefont {C.~P.}\ \bibnamefont {Moca}}, \bibinfo {author} {\bibfnamefont {D.}~\bibnamefont {Sticlet}}, \bibinfo {author} {\bibfnamefont {B.}~\bibnamefont {D{\'o}ra}}, \bibinfo {author} {\bibfnamefont {A.}~\bibnamefont {Valli}}, \bibinfo {author} {\bibfnamefont {D.}~\bibnamefont {Szombathy}},\ and\ \bibinfo {author} {\bibfnamefont {G.}~\bibnamefont {Zar{\'a}nd}},\ }\bibfield  {title} {\bibinfo {title} {Non-stabilizerness generation in a multi-particle quantum walk},\ }\bibfield  {journal} {\bibinfo  {journal} {arXiv preprint arXiv:2504.19750}\ }\href {https://doi.org/10.48550/arXiv.2504.19750} {10.48550/arXiv.2504.19750} (\bibinfo {year} {2025})\BibitemShut {NoStop}%
\bibitem [{\citenamefont {Tarabunga}\ and\ \citenamefont {Haug}(2025)}]{tarabunga2025efficient}%
  \BibitemOpen
  \bibfield  {author} {\bibinfo {author} {\bibfnamefont {P.~S.}\ \bibnamefont {Tarabunga}}\ and\ \bibinfo {author} {\bibfnamefont {T.}~\bibnamefont {Haug}},\ }\bibfield  {title} {\bibinfo {title} {Efficient mutual magic and magic capacity with matrix product states},\ }\href {https://doi.org/10.21468/SciPostPhys.19.4.085} {\bibfield  {journal} {\bibinfo  {journal} {SciPost Phys.}\ }\textbf {\bibinfo {volume} {19}},\ \bibinfo {pages} {085} (\bibinfo {year} {2025})}\BibitemShut {NoStop}%
\bibitem [{\citenamefont {Jasser}\ \emph {et~al.}(2025)\citenamefont {Jasser}, \citenamefont {Odavi{\'c}},\ and\ \citenamefont {Hamma}}]{jasser2025stabilizer}%
  \BibitemOpen
  \bibfield  {author} {\bibinfo {author} {\bibfnamefont {B.}~\bibnamefont {Jasser}}, \bibinfo {author} {\bibfnamefont {J.}~\bibnamefont {Odavi{\'c}}},\ and\ \bibinfo {author} {\bibfnamefont {A.}~\bibnamefont {Hamma}},\ }\bibfield  {title} {\bibinfo {title} {Stabilizer entropy and entanglement complexity in the sachdev-ye-kitaev model},\ }\href {https://doi.org/10.1103/rz86-47h3} {\bibfield  {journal} {\bibinfo  {journal} {Physical Review B}\ }\textbf {\bibinfo {volume} {112}},\ \bibinfo {pages} {174204} (\bibinfo {year} {2025})}\BibitemShut {NoStop}%
\bibitem [{\citenamefont {Cianciulli}\ \emph {et~al.}(2024)\citenamefont {Cianciulli}, \citenamefont {Rossignoli}, \citenamefont {Di~Tullio}, \citenamefont {Gigena},\ and\ \citenamefont {Petrovich}}]{cianciulli2024bipartite}%
  \BibitemOpen
  \bibfield  {author} {\bibinfo {author} {\bibfnamefont {J.}~\bibnamefont {Cianciulli}}, \bibinfo {author} {\bibfnamefont {R.}~\bibnamefont {Rossignoli}}, \bibinfo {author} {\bibfnamefont {M.}~\bibnamefont {Di~Tullio}}, \bibinfo {author} {\bibfnamefont {N.}~\bibnamefont {Gigena}},\ and\ \bibinfo {author} {\bibfnamefont {F.}~\bibnamefont {Petrovich}},\ }\bibfield  {title} {\bibinfo {title} {Bipartite representations and many-body entanglement of pure states of n indistinguishable particles},\ }\href {https://doi.org/10.1103/PhysRevA.110.032414} {\bibfield  {journal} {\bibinfo  {journal} {Physical Review A}\ }\textbf {\bibinfo {volume} {110}},\ \bibinfo {pages} {032414} (\bibinfo {year} {2024})}\BibitemShut {NoStop}%
\bibitem [{\citenamefont {Gigena}\ \emph {et~al.}(2021)\citenamefont {Gigena}, \citenamefont {Di~Tullio},\ and\ \citenamefont {Rossignoli}}]{gigena2021many}%
  \BibitemOpen
  \bibfield  {author} {\bibinfo {author} {\bibfnamefont {N.}~\bibnamefont {Gigena}}, \bibinfo {author} {\bibfnamefont {M.}~\bibnamefont {Di~Tullio}},\ and\ \bibinfo {author} {\bibfnamefont {R.}~\bibnamefont {Rossignoli}},\ }\bibfield  {title} {\bibinfo {title} {Many-body entanglement in fermion systems},\ }\href {https://doi.org/10.1103/PhysRevA.103.052424} {\bibfield  {journal} {\bibinfo  {journal} {Physical Review A}\ }\textbf {\bibinfo {volume} {103}},\ \bibinfo {pages} {052424} (\bibinfo {year} {2021})}\BibitemShut {NoStop}%
\bibitem [{\citenamefont {Collura}\ \emph {et~al.}(2024)\citenamefont {Collura}, \citenamefont {De~Nardis}, \citenamefont {Alba},\ and\ \citenamefont {Lami}}]{collura2024quantum}%
  \BibitemOpen
  \bibfield  {author} {\bibinfo {author} {\bibfnamefont {M.}~\bibnamefont {Collura}}, \bibinfo {author} {\bibfnamefont {J.}~\bibnamefont {De~Nardis}}, \bibinfo {author} {\bibfnamefont {V.}~\bibnamefont {Alba}},\ and\ \bibinfo {author} {\bibfnamefont {G.}~\bibnamefont {Lami}},\ }\bibfield  {title} {\bibinfo {title} {The quantum magic of fermionic gaussian states},\ }\bibfield  {journal} {\bibinfo  {journal} {arXiv preprint arXiv:2412.05367}\ }\href {https://doi.org/10.48550/arXiv.2412.05367} {10.48550/arXiv.2412.05367} (\bibinfo {year} {2024})\BibitemShut {NoStop}%
\bibitem [{\citenamefont {Gu}\ \emph {et~al.}(2025)\citenamefont {Gu}, \citenamefont {Oliviero},\ and\ \citenamefont {Leone}}]{gu2025magic}%
  \BibitemOpen
  \bibfield  {author} {\bibinfo {author} {\bibfnamefont {A.}~\bibnamefont {Gu}}, \bibinfo {author} {\bibfnamefont {S.~F.}\ \bibnamefont {Oliviero}},\ and\ \bibinfo {author} {\bibfnamefont {L.}~\bibnamefont {Leone}},\ }\bibfield  {title} {\bibinfo {title} {Magic-induced computational separation in entanglement theory},\ }\href {https://doi.org/10.1103/PRXQuantum.6.020324} {\bibfield  {journal} {\bibinfo  {journal} {PRX Quantum}\ }\textbf {\bibinfo {volume} {6}},\ \bibinfo {pages} {020324} (\bibinfo {year} {2025})}\BibitemShut {NoStop}%
\bibitem [{\citenamefont {Koma}\ and\ \citenamefont {Tasaki}(1994)}]{koma1994symmetry}%
  \BibitemOpen
  \bibfield  {author} {\bibinfo {author} {\bibfnamefont {T.}~\bibnamefont {Koma}}\ and\ \bibinfo {author} {\bibfnamefont {H.}~\bibnamefont {Tasaki}},\ }\bibfield  {title} {\bibinfo {title} {Symmetry breaking and finite-size effects in quantum many-body systems},\ }\href {https://doi.org/10.1007/BF02188685} {\bibfield  {journal} {\bibinfo  {journal} {Journal of statistical physics}\ }\textbf {\bibinfo {volume} {76}},\ \bibinfo {pages} {745} (\bibinfo {year} {1994})}\BibitemShut {NoStop}%
\bibitem [{\citenamefont {Okatev}\ \emph {et~al.}(2025)\citenamefont {Okatev}, \citenamefont {Sotnikov},\ and\ \citenamefont {Mazurenko}}]{okatev2025exploring}%
  \BibitemOpen
  \bibfield  {author} {\bibinfo {author} {\bibfnamefont {V.}~\bibnamefont {Okatev}}, \bibinfo {author} {\bibfnamefont {O.~M.}\ \bibnamefont {Sotnikov}},\ and\ \bibinfo {author} {\bibfnamefont {V.}~\bibnamefont {Mazurenko}},\ }\bibfield  {title} {\bibinfo {title} {Exploring entanglement in finite-size quantum systems with degenerate ground state},\ }\href {https://doi.org/10.1103/PhysRevB.111.054443} {\bibfield  {journal} {\bibinfo  {journal} {Physical Review B}\ }\textbf {\bibinfo {volume} {111}},\ \bibinfo {pages} {054443} (\bibinfo {year} {2025})}\BibitemShut {NoStop}%
\bibitem [{\citenamefont {Cerezo}\ \emph {et~al.}(2015)\citenamefont {Cerezo}, \citenamefont {Rossignoli},\ and\ \citenamefont {Canosa}}]{cerezo2015nontransverse}%
  \BibitemOpen
  \bibfield  {author} {\bibinfo {author} {\bibfnamefont {M.}~\bibnamefont {Cerezo}}, \bibinfo {author} {\bibfnamefont {R.}~\bibnamefont {Rossignoli}},\ and\ \bibinfo {author} {\bibfnamefont {N.}~\bibnamefont {Canosa}},\ }\bibfield  {title} {\bibinfo {title} {Nontransverse factorizing fields and entanglement in finite spin systems},\ }\href {https://doi.org/10.1103/PhysRevB.92.224422} {\bibfield  {journal} {\bibinfo  {journal} {Physical Review B}\ }\textbf {\bibinfo {volume} {92}},\ \bibinfo {pages} {224422} (\bibinfo {year} {2015})}\BibitemShut {NoStop}%
\bibitem [{\citenamefont {Cerezo}\ \emph {et~al.}(2017)\citenamefont {Cerezo}, \citenamefont {Rossignoli}, \citenamefont {Canosa},\ and\ \citenamefont {R{\'\i}os}}]{cerezo2017factorization}%
  \BibitemOpen
  \bibfield  {author} {\bibinfo {author} {\bibfnamefont {M.}~\bibnamefont {Cerezo}}, \bibinfo {author} {\bibfnamefont {R.}~\bibnamefont {Rossignoli}}, \bibinfo {author} {\bibfnamefont {N.}~\bibnamefont {Canosa}},\ and\ \bibinfo {author} {\bibfnamefont {E.}~\bibnamefont {R{\'\i}os}},\ }\bibfield  {title} {\bibinfo {title} {Factorization and criticality in finite $xxz$ systems of arbitrary spin},\ }\href {https://doi.org/10.1103/PhysRevLett.119.220605} {\bibfield  {journal} {\bibinfo  {journal} {Physical Review Letters}\ }\textbf {\bibinfo {volume} {119}},\ \bibinfo {pages} {220605} (\bibinfo {year} {2017})}\BibitemShut {NoStop}%
\bibitem [{\citenamefont {Petrovich}\ \emph {et~al.}(2022)\citenamefont {Petrovich}, \citenamefont {Canosa},\ and\ \citenamefont {Rossignoli}}]{petrovich2022ground}%
  \BibitemOpen
  \bibfield  {author} {\bibinfo {author} {\bibfnamefont {F.}~\bibnamefont {Petrovich}}, \bibinfo {author} {\bibfnamefont {N.}~\bibnamefont {Canosa}},\ and\ \bibinfo {author} {\bibfnamefont {R.}~\bibnamefont {Rossignoli}},\ }\bibfield  {title} {\bibinfo {title} {Ground-state separability and criticality in interacting many-particle systems},\ }\href {https://doi.org/10.1103/PhysRevA.105.032212} {\bibfield  {journal} {\bibinfo  {journal} {Physical Review A}\ }\textbf {\bibinfo {volume} {105}},\ \bibinfo {pages} {032212} (\bibinfo {year} {2022})}\BibitemShut {NoStop}%
\bibitem [{\citenamefont {Wilde}(2013)}]{wilde2013quantum}%
  \BibitemOpen
  \bibfield  {author} {\bibinfo {author} {\bibfnamefont {M.~M.}\ \bibnamefont {Wilde}},\ }\href@noop {} {\emph {\bibinfo {title} {Quantum Information Theory}}}\ (\bibinfo  {publisher} {Cambridge University Press},\ \bibinfo {address} {Cambridge, UK},\ \bibinfo {year} {2013})\BibitemShut {NoStop}%
\bibitem [{\citenamefont {Horodecki}\ \emph {et~al.}(2009)\citenamefont {Horodecki}, \citenamefont {Horodecki}, \citenamefont {Horodecki},\ and\ \citenamefont {Horodecki}}]{horodecki2009quantum}%
  \BibitemOpen
  \bibfield  {author} {\bibinfo {author} {\bibfnamefont {R.}~\bibnamefont {Horodecki}}, \bibinfo {author} {\bibfnamefont {P.}~\bibnamefont {Horodecki}}, \bibinfo {author} {\bibfnamefont {M.}~\bibnamefont {Horodecki}},\ and\ \bibinfo {author} {\bibfnamefont {K.}~\bibnamefont {Horodecki}},\ }\bibfield  {title} {\bibinfo {title} {Quantum entanglement},\ }\href {https://doi.org/10.1103/RevModPhys.81.865} {\bibfield  {journal} {\bibinfo  {journal} {Reviews of modern physics}\ }\textbf {\bibinfo {volume} {81}},\ \bibinfo {pages} {865} (\bibinfo {year} {2009})}\BibitemShut {NoStop}%
\bibitem [{\citenamefont {Schliemann}\ \emph {et~al.}(2001)\citenamefont {Schliemann}, \citenamefont {Cirac}, \citenamefont {Kus}, \citenamefont {Lewenstein},\ and\ \citenamefont {Loss}}]{schliemann2001quantum}%
  \BibitemOpen
  \bibfield  {author} {\bibinfo {author} {\bibfnamefont {J.}~\bibnamefont {Schliemann}}, \bibinfo {author} {\bibfnamefont {J.~I.}\ \bibnamefont {Cirac}}, \bibinfo {author} {\bibfnamefont {M.}~\bibnamefont {Kus}}, \bibinfo {author} {\bibfnamefont {M.}~\bibnamefont {Lewenstein}},\ and\ \bibinfo {author} {\bibfnamefont {D.}~\bibnamefont {Loss}},\ }\bibfield  {title} {\bibinfo {title} {Quantum correlations in two-fermion systems},\ }\href {https://doi.org/10.1103/PhysRevA.64.022303} {\bibfield  {journal} {\bibinfo  {journal} {Physical Review A}\ }\textbf {\bibinfo {volume} {64}},\ \bibinfo {pages} {022303} (\bibinfo {year} {2001})}\BibitemShut {NoStop}%
\bibitem [{\citenamefont {Gigena}\ and\ \citenamefont {Rossignoli}(2015)}]{gigena2015entanglement}%
  \BibitemOpen
  \bibfield  {author} {\bibinfo {author} {\bibfnamefont {N.}~\bibnamefont {Gigena}}\ and\ \bibinfo {author} {\bibfnamefont {R.}~\bibnamefont {Rossignoli}},\ }\bibfield  {title} {\bibinfo {title} {Entanglement in fermion systems},\ }\href {https://doi.org/10.1103/PhysRevA.92.042326} {\bibfield  {journal} {\bibinfo  {journal} {Physical Review A}\ }\textbf {\bibinfo {volume} {92}},\ \bibinfo {pages} {042326} (\bibinfo {year} {2015})}\BibitemShut {NoStop}%
\bibitem [{\citenamefont {Gigena}\ and\ \citenamefont {Rossignoli}(2017)}]{gigena2017bipartite}%
  \BibitemOpen
  \bibfield  {author} {\bibinfo {author} {\bibfnamefont {N.}~\bibnamefont {Gigena}}\ and\ \bibinfo {author} {\bibfnamefont {R.}~\bibnamefont {Rossignoli}},\ }\bibfield  {title} {\bibinfo {title} {Bipartite entanglement in fermion systems},\ }\href {https://doi.org/10.1103/PhysRevA.95.062320} {\bibfield  {journal} {\bibinfo  {journal} {Physical Review A}\ }\textbf {\bibinfo {volume} {95}},\ \bibinfo {pages} {062320} (\bibinfo {year} {2017})}\BibitemShut {NoStop}%
\bibitem [{\citenamefont {Di~Tullio}\ \emph {et~al.}(2019)\citenamefont {Di~Tullio}, \citenamefont {Rossignoli}, \citenamefont {Cerezo},\ and\ \citenamefont {Gigena}}]{di2019fermionic}%
  \BibitemOpen
  \bibfield  {author} {\bibinfo {author} {\bibfnamefont {M.}~\bibnamefont {Di~Tullio}}, \bibinfo {author} {\bibfnamefont {R.}~\bibnamefont {Rossignoli}}, \bibinfo {author} {\bibfnamefont {M.}~\bibnamefont {Cerezo}},\ and\ \bibinfo {author} {\bibfnamefont {N.}~\bibnamefont {Gigena}},\ }\bibfield  {title} {\bibinfo {title} {Fermionic entanglement in the {L}ipkin model},\ }\href {https://doi.org/10.1103/PhysRevA.100.062104} {\bibfield  {journal} {\bibinfo  {journal} {Physical Review A}\ }\textbf {\bibinfo {volume} {100}},\ \bibinfo {pages} {062104} (\bibinfo {year} {2019})}\BibitemShut {NoStop}%
\bibitem [{\citenamefont {Gigena}\ \emph {et~al.}(2020)\citenamefont {Gigena}, \citenamefont {Di~Tullio},\ and\ \citenamefont {Rossignoli}}]{gigena2020one}%
  \BibitemOpen
  \bibfield  {author} {\bibinfo {author} {\bibfnamefont {N.}~\bibnamefont {Gigena}}, \bibinfo {author} {\bibfnamefont {M.}~\bibnamefont {Di~Tullio}},\ and\ \bibinfo {author} {\bibfnamefont {R.}~\bibnamefont {Rossignoli}},\ }\bibfield  {title} {\bibinfo {title} {One-body entanglement as a quantum resource in fermionic systems},\ }\href {https://doi.org/10.1103/PhysRevA.102.042410} {\bibfield  {journal} {\bibinfo  {journal} {Physical Review A}\ }\textbf {\bibinfo {volume} {102}},\ \bibinfo {pages} {042410} (\bibinfo {year} {2020})}\BibitemShut {NoStop}%
\bibitem [{\citenamefont {Benatti}\ \emph {et~al.}(2020)\citenamefont {Benatti}, \citenamefont {Floreanini}, \citenamefont {Franchini},\ and\ \citenamefont {Marzolino}}]{benatti2020entanglement}%
  \BibitemOpen
  \bibfield  {author} {\bibinfo {author} {\bibfnamefont {F.}~\bibnamefont {Benatti}}, \bibinfo {author} {\bibfnamefont {R.}~\bibnamefont {Floreanini}}, \bibinfo {author} {\bibfnamefont {F.}~\bibnamefont {Franchini}},\ and\ \bibinfo {author} {\bibfnamefont {U.}~\bibnamefont {Marzolino}},\ }\bibfield  {title} {\bibinfo {title} {Entanglement in indistinguishable particle systems},\ }\href {https://doi.org/10.1016/j.physrep.2020.07.003} {\bibfield  {journal} {\bibinfo  {journal} {Physics Reports}\ }\textbf {\bibinfo {volume} {878}},\ \bibinfo {pages} {1} (\bibinfo {year} {2020})}\BibitemShut {NoStop}%
\bibitem [{\citenamefont {Garcia}\ \emph {et~al.}(2026)\citenamefont {Garcia}, \citenamefont {Cianciulli},\ and\ \citenamefont {Rossignoli}}]{garcia2026fermionic}%
  \BibitemOpen
  \bibfield  {author} {\bibinfo {author} {\bibfnamefont {J.}~\bibnamefont {Garcia}}, \bibinfo {author} {\bibfnamefont {J.}~\bibnamefont {Cianciulli}},\ and\ \bibinfo {author} {\bibfnamefont {R.}~\bibnamefont {Rossignoli}},\ }\bibfield  {title} {\bibinfo {title} {Fermionic entanglement and quantum correlation measures in molecules},\ }\bibfield  {journal} {\bibinfo  {journal} {arXiv preprint arXiv:2604.07633}\ }\href {https://doi.org/10.48550/arXiv.2604.07633} {10.48550/arXiv.2604.07633} (\bibinfo {year} {2026})\BibitemShut {NoStop}%
\bibitem [{\citenamefont {Bittel}\ \emph {et~al.}(2025)\citenamefont {Bittel}, \citenamefont {Mele}, \citenamefont {Eisert},\ and\ \citenamefont {Leone}}]{bittel2025optimal}%
  \BibitemOpen
  \bibfield  {author} {\bibinfo {author} {\bibfnamefont {L.}~\bibnamefont {Bittel}}, \bibinfo {author} {\bibfnamefont {A.~A.}\ \bibnamefont {Mele}}, \bibinfo {author} {\bibfnamefont {J.}~\bibnamefont {Eisert}},\ and\ \bibinfo {author} {\bibfnamefont {L.}~\bibnamefont {Leone}},\ }\bibfield  {title} {\bibinfo {title} {Optimal trace-distance bounds for free-fermionic states: Testing and improved tomography},\ }\href {https://doi.org/10.1103/pzx6-nkfb} {\bibfield  {journal} {\bibinfo  {journal} {PRX Quantum}\ }\textbf {\bibinfo {volume} {6}},\ \bibinfo {pages} {030341} (\bibinfo {year} {2025})}\BibitemShut {NoStop}%
\bibitem [{\citenamefont {Walter}\ and\ \citenamefont {Witteveen}(2026)}]{tarabunga2026computable}%
  \BibitemOpen
  \bibfield  {author} {\bibinfo {author} {\bibfnamefont {M.}~\bibnamefont {Walter}}\ and\ \bibinfo {author} {\bibfnamefont {F.}~\bibnamefont {Witteveen}},\ }\bibfield  {title} {\bibinfo {title} {Computable measures of fermionic non-{Gaussianity} from the covariance matrix},\ }\bibfield  {journal} {\bibinfo  {journal} {arXiv preprint arXiv:2607.02242}\ }\href {https://doi.org/10.48550/arXiv.2607.02242} {10.48550/arXiv.2607.02242} (\bibinfo {year} {2026})\BibitemShut {NoStop}%
\bibitem [{\citenamefont {Braccia}\ \emph {et~al.}(2026)\citenamefont {Braccia}, \citenamefont {Diaz}, \citenamefont {Larocca}, \citenamefont {Cerezo},\ and\ \citenamefont {Garc{\'\i}a-Mart{\'\i}n}}]{braccia2026commutant}%
  \BibitemOpen
  \bibfield  {author} {\bibinfo {author} {\bibfnamefont {P.}~\bibnamefont {Braccia}}, \bibinfo {author} {\bibfnamefont {N.}~\bibnamefont {Diaz}}, \bibinfo {author} {\bibfnamefont {M.}~\bibnamefont {Larocca}}, \bibinfo {author} {\bibfnamefont {M.}~\bibnamefont {Cerezo}},\ and\ \bibinfo {author} {\bibfnamefont {D.}~\bibnamefont {Garc{\'\i}a-Mart{\'\i}n}},\ }\bibfield  {title} {\bibinfo {title} {The commutant of fermionic gaussian unitaries},\ }\bibfield  {journal} {\bibinfo  {journal} {arXiv preprint arXiv:2603.19210}\ }\href {https://doi.org/10.48550/arXiv.2603.19210} {10.48550/arXiv.2603.19210} (\bibinfo {year} {2026})\BibitemShut {NoStop}%
\bibitem [{\citenamefont {Sierant}\ \emph {et~al.}(2026)\citenamefont {Sierant}, \citenamefont {X.},\ and\ \citenamefont {Tarabunga}}]{sierant2026theory}%
  \BibitemOpen
  \bibfield  {author} {\bibinfo {author} {\bibfnamefont {P.}~\bibnamefont {Sierant}}, \bibinfo {author} {\bibfnamefont {T.}~\bibnamefont {X.}},\ and\ \bibinfo {author} {\bibfnamefont {P.~S.}\ \bibnamefont {Tarabunga}},\ }\bibfield  {title} {\bibinfo {title} {Theory of the matchgate commutant},\ }\bibfield  {journal} {\bibinfo  {journal} {arXiv preprint arXiv:2603.12392}\ }\href {https://doi.org/10.48550/arXiv.2603.12392} {10.48550/arXiv.2603.12392} (\bibinfo {year} {2026})\BibitemShut {NoStop}%
\bibitem [{\citenamefont {West}\ \emph {et~al.}(2026)\citenamefont {West}, \citenamefont {Chang}, \citenamefont {Coffman}, \citenamefont {Larocca},\ and\ \citenamefont {Cerezo}}]{west2026fermionic}%
  \BibitemOpen
  \bibfield  {author} {\bibinfo {author} {\bibfnamefont {M.}~\bibnamefont {West}}, \bibinfo {author} {\bibfnamefont {S.~Y.}\ \bibnamefont {Chang}}, \bibinfo {author} {\bibfnamefont {L.}~\bibnamefont {Coffman}}, \bibinfo {author} {\bibfnamefont {M.}~\bibnamefont {Larocca}},\ and\ \bibinfo {author} {\bibfnamefont {M.}~\bibnamefont {Cerezo}},\ }\bibfield  {title} {\bibinfo {title} {Practical fermionic shadows enabled by improved sample-complexity bounds},\ }\bibfield  {journal} {\bibinfo  {journal} {arXiv preprint arXiv:2609.40173}\ }\href {https://doi.org/10.48550/arXiv.2609.40173} {10.48550/arXiv.2609.40173} (\bibinfo {year} {2026})\BibitemShut {NoStop}%
\bibitem [{\citenamefont {Petrovich}\ \emph {et~al.}(2026)\citenamefont {Petrovich}, \citenamefont {Rossignoli},\ and\ \citenamefont {Cerezo}}]{petrovich2026symmetry}%
  \BibitemOpen
  \bibfield  {author} {\bibinfo {author} {\bibfnamefont {F.}~\bibnamefont {Petrovich}}, \bibinfo {author} {\bibfnamefont {R.}~\bibnamefont {Rossignoli}},\ and\ \bibinfo {author} {\bibfnamefont {M.}~\bibnamefont {Cerezo}},\ }\bibfield  {title} {\bibinfo {title} {Classification of quantum states through conserved operators},\ }\bibfield  {journal} {\bibinfo  {journal} {arXiv preprint arXiv:2610.20988}\ }\href {https://doi.org/10.48550/arXiv.2610.20988} {10.48550/arXiv.2610.20988} (\bibinfo {year} {2026})\BibitemShut {NoStop}%
\bibitem [{\citenamefont {Staros}\ \emph {et~al.}(2026)\citenamefont {Staros}, \citenamefont {Petrovich}, \citenamefont {Kirmani}, \citenamefont {Blaschke}, \citenamefont {Cerezo},\ and\ \citenamefont {Zhu}}]{staros2026towards}%
  \BibitemOpen
  \bibfield  {author} {\bibinfo {author} {\bibfnamefont {D.}~\bibnamefont {Staros}}, \bibinfo {author} {\bibfnamefont {F.}~\bibnamefont {Petrovich}}, \bibinfo {author} {\bibfnamefont {A.}~\bibnamefont {Kirmani}}, \bibinfo {author} {\bibfnamefont {D.}~\bibnamefont {Blaschke}}, \bibinfo {author} {\bibfnamefont {M.}~\bibnamefont {Cerezo}},\ and\ \bibinfo {author} {\bibfnamefont {J.-X.}\ \bibnamefont {Zhu}},\ }\bibfield  {title} {\bibinfo {title} {Towards quantum computation of bulk feo},\ }\href@noop {} {\bibfield  {journal} {\bibinfo  {journal} {Manuscript in preparation}\ } (\bibinfo {year} {2026})}\BibitemShut {NoStop}%
\bibitem [{\citenamefont {Nielsen}\ and\ \citenamefont {Chuang}(2000)}]{nielsen2000quantum}%
  \BibitemOpen
  \bibfield  {author} {\bibinfo {author} {\bibfnamefont {M.~A.}\ \bibnamefont {Nielsen}}\ and\ \bibinfo {author} {\bibfnamefont {I.~L.}\ \bibnamefont {Chuang}},\ }\href@noop {} {\emph {\bibinfo {title} {Quantum Computation and Quantum Information}}}\ (\bibinfo  {publisher} {Cambridge University Press},\ \bibinfo {address} {Cambridge},\ \bibinfo {year} {2000})\BibitemShut {NoStop}%
\bibitem [{\citenamefont {Diaz}\ \emph {et~al.}(2023)\citenamefont {Diaz}, \citenamefont {Garc{\'\i}a-Mart{\'\i}n}, \citenamefont {Kazi}, \citenamefont {Larocca},\ and\ \citenamefont {Cerezo}}]{diaz2023showcasing}%
  \BibitemOpen
  \bibfield  {author} {\bibinfo {author} {\bibfnamefont {N.~L.}\ \bibnamefont {Diaz}}, \bibinfo {author} {\bibfnamefont {D.}~\bibnamefont {Garc{\'\i}a-Mart{\'\i}n}}, \bibinfo {author} {\bibfnamefont {S.}~\bibnamefont {Kazi}}, \bibinfo {author} {\bibfnamefont {M.}~\bibnamefont {Larocca}},\ and\ \bibinfo {author} {\bibfnamefont {M.}~\bibnamefont {Cerezo}},\ }\bibfield  {title} {\bibinfo {title} {Showcasing a barren plateau theory beyond the dynamical lie algebra},\ }\bibfield  {journal} {\bibinfo  {journal} {arXiv preprint arXiv:2310.11505}\ }\href {https://doi.org/10.48550/arXiv.2310.11505} {10.48550/arXiv.2310.11505} (\bibinfo {year} {2023})\BibitemShut {NoStop}%
\bibitem [{\citenamefont {Chitambar}\ and\ \citenamefont {Gour}(2019)}]{chitambar2019quantum}%
  \BibitemOpen
  \bibfield  {author} {\bibinfo {author} {\bibfnamefont {E.}~\bibnamefont {Chitambar}}\ and\ \bibinfo {author} {\bibfnamefont {G.}~\bibnamefont {Gour}},\ }\bibfield  {title} {\bibinfo {title} {Quantum resource theories},\ }\href {https://doi.org/10.1103/RevModPhys.91.025001} {\bibfield  {journal} {\bibinfo  {journal} {Reviews of modern physics}\ }\textbf {\bibinfo {volume} {91}},\ \bibinfo {pages} {025001} (\bibinfo {year} {2019})}\BibitemShut {NoStop}%
\bibitem [{\citenamefont {Brandao}\ and\ \citenamefont {Gour}(2015)}]{brandao2015reversible}%
  \BibitemOpen
  \bibfield  {author} {\bibinfo {author} {\bibfnamefont {F.~G.}\ \bibnamefont {Brandao}}\ and\ \bibinfo {author} {\bibfnamefont {G.}~\bibnamefont {Gour}},\ }\bibfield  {title} {\bibinfo {title} {Reversible framework for quantum resource theories},\ }\href {https://doi.org/10.1103/PhysRevLett.115.070503} {\bibfield  {journal} {\bibinfo  {journal} {Physical Review Letters}\ }\textbf {\bibinfo {volume} {115}},\ \bibinfo {pages} {070503} (\bibinfo {year} {2015})}\BibitemShut {NoStop}%
\bibitem [{\citenamefont {Chitambar}\ \emph {et~al.}(2014)\citenamefont {Chitambar}, \citenamefont {Leung}, \citenamefont {Man{\v{c}}inska}, \citenamefont {Ozols},\ and\ \citenamefont {Winter}}]{chitambar2014everything}%
  \BibitemOpen
  \bibfield  {author} {\bibinfo {author} {\bibfnamefont {E.}~\bibnamefont {Chitambar}}, \bibinfo {author} {\bibfnamefont {D.}~\bibnamefont {Leung}}, \bibinfo {author} {\bibfnamefont {L.}~\bibnamefont {Man{\v{c}}inska}}, \bibinfo {author} {\bibfnamefont {M.}~\bibnamefont {Ozols}},\ and\ \bibinfo {author} {\bibfnamefont {A.}~\bibnamefont {Winter}},\ }\bibfield  {title} {\bibinfo {title} {Everything you always wanted to know about locc (but were afraid to ask)},\ }\href {https://doi.org/10.1007/s00220-014-1953-9} {\bibfield  {journal} {\bibinfo  {journal} {Communications in Mathematical Physics}\ }\textbf {\bibinfo {volume} {328}},\ \bibinfo {pages} {303} (\bibinfo {year} {2014})}\BibitemShut {NoStop}%
\bibitem [{\citenamefont {D{\"u}r}\ \emph {et~al.}(2000)\citenamefont {D{\"u}r}, \citenamefont {Vidal},\ and\ \citenamefont {Cirac}}]{dur2000three}%
  \BibitemOpen
  \bibfield  {author} {\bibinfo {author} {\bibfnamefont {W.}~\bibnamefont {D{\"u}r}}, \bibinfo {author} {\bibfnamefont {G.}~\bibnamefont {Vidal}},\ and\ \bibinfo {author} {\bibfnamefont {J.~I.}\ \bibnamefont {Cirac}},\ }\bibfield  {title} {\bibinfo {title} {Three qubits can be entangled in two inequivalent ways},\ }\href {https://doi.org/10.1103/PhysRevA.62.062314} {\bibfield  {journal} {\bibinfo  {journal} {Physical Review A}\ }\textbf {\bibinfo {volume} {62}},\ \bibinfo {pages} {062314} (\bibinfo {year} {2000})}\BibitemShut {NoStop}%
\bibitem [{\citenamefont {Miyake}(2003)}]{miyake2003classification}%
  \BibitemOpen
  \bibfield  {author} {\bibinfo {author} {\bibfnamefont {A.}~\bibnamefont {Miyake}},\ }\bibfield  {title} {\bibinfo {title} {Classification of multipartite entangled states by multidimensional determinants},\ }\href {https://doi.org/10.1103/PhysRevA.67.012108} {\bibfield  {journal} {\bibinfo  {journal} {Physical Review A}\ }\textbf {\bibinfo {volume} {67}},\ \bibinfo {pages} {012108} (\bibinfo {year} {2003})}\BibitemShut {NoStop}%
\bibitem [{\citenamefont {Verstraete}\ \emph {et~al.}(2002)\citenamefont {Verstraete}, \citenamefont {Dehaene}, \citenamefont {De~Moor},\ and\ \citenamefont {Verschelde}}]{verstraete2002four}%
  \BibitemOpen
  \bibfield  {author} {\bibinfo {author} {\bibfnamefont {F.}~\bibnamefont {Verstraete}}, \bibinfo {author} {\bibfnamefont {J.}~\bibnamefont {Dehaene}}, \bibinfo {author} {\bibfnamefont {B.}~\bibnamefont {De~Moor}},\ and\ \bibinfo {author} {\bibfnamefont {H.}~\bibnamefont {Verschelde}},\ }\bibfield  {title} {\bibinfo {title} {Four qubits can be entangled in nine different ways},\ }\href {https://doi.org/10.1103/PhysRevA.65.052112} {\bibfield  {journal} {\bibinfo  {journal} {Physical Review A}\ }\textbf {\bibinfo {volume} {65}},\ \bibinfo {pages} {052112} (\bibinfo {year} {2002})}\BibitemShut {NoStop}%
\bibitem [{\citenamefont {Dietrich}\ \emph {et~al.}(2022)\citenamefont {Dietrich}, \citenamefont {De~Graaf}, \citenamefont {Marrani},\ and\ \citenamefont {Origlia}}]{dietrich2022classification}%
  \BibitemOpen
  \bibfield  {author} {\bibinfo {author} {\bibfnamefont {H.}~\bibnamefont {Dietrich}}, \bibinfo {author} {\bibfnamefont {W.~A.}\ \bibnamefont {De~Graaf}}, \bibinfo {author} {\bibfnamefont {A.}~\bibnamefont {Marrani}},\ and\ \bibinfo {author} {\bibfnamefont {M.}~\bibnamefont {Origlia}},\ }\bibfield  {title} {\bibinfo {title} {Classification of four qubit states and their stabilisers under slocc operations},\ }\href {https://doi.org/10.1088/1751-8121/ac4b13} {\bibfield  {journal} {\bibinfo  {journal} {Journal of Physics A: Mathematical and Theoretical}\ }\textbf {\bibinfo {volume} {55}},\ \bibinfo {pages} {095302} (\bibinfo {year} {2022})}\BibitemShut {NoStop}%
\bibitem [{\citenamefont {Jia}\ \emph {et~al.}(2015)\citenamefont {Jia}, \citenamefont {Li},\ and\ \citenamefont {Qiao}}]{jia2015classification}%
  \BibitemOpen
  \bibfield  {author} {\bibinfo {author} {\bibfnamefont {K.-K.}\ \bibnamefont {Jia}}, \bibinfo {author} {\bibfnamefont {J.-L.}\ \bibnamefont {Li}},\ and\ \bibinfo {author} {\bibfnamefont {C.-F.}\ \bibnamefont {Qiao}},\ }\bibfield  {title} {\bibinfo {title} {Classification of the entangled states of $2\times l\times m\times n\times h$},\ }\bibfield  {journal} {\bibinfo  {journal} {arXiv preprint arXiv:1505.02007}\ }\href {https://doi.org/10.48550/arXiv.1505.02007} {10.48550/arXiv.1505.02007} (\bibinfo {year} {2015})\BibitemShut {NoStop}%
\bibitem [{\citenamefont {Diaz}\ \emph {et~al.}(2025)\citenamefont {Diaz}, \citenamefont {Mele}, \citenamefont {Bermejo}, \citenamefont {Braccia}, \citenamefont {Larocca},\ and\ \citenamefont {Cerezo}}]{diaz2025unified}%
  \BibitemOpen
  \bibfield  {author} {\bibinfo {author} {\bibfnamefont {N.~L.}\ \bibnamefont {Diaz}}, \bibinfo {author} {\bibfnamefont {A.~A.}\ \bibnamefont {Mele}}, \bibinfo {author} {\bibfnamefont {P.}~\bibnamefont {Bermejo}}, \bibinfo {author} {\bibfnamefont {P.}~\bibnamefont {Braccia}}, \bibinfo {author} {\bibfnamefont {M.}~\bibnamefont {Larocca}},\ and\ \bibinfo {author} {\bibfnamefont {M.}~\bibnamefont {Cerezo}},\ }\bibfield  {title} {\bibinfo {title} {A unified approach to quantum resource theories and a new class of free operations},\ }\bibfield  {journal} {\bibinfo  {journal} {arXiv preprint arXiv:2507.10851}\ }\href {https://doi.org/10.48550/arXiv.2507.10851} {10.48550/arXiv.2507.10851} (\bibinfo {year} {2025})\BibitemShut {NoStop}%
\end{thebibliography}%

\clearpage
\newpage
\appendix

\section{One-body Hamiltonian of FeO}
\label{app:FeO}

Here we provide the one-body matrices
${\bf h}^{\pi_y\pi_z}$ of Eq.~\eqref{hpiypiz}.
In the orbital bases defined by $\mathcal C_{\pi_y\pi_z}$, they read

\begin{align}
{\bf h}^{++}
&=
\begin{pmatrix}
-8.1 & -1.0 & 1.8\\
-1.0 & -5.3 & 0.66\\
1.8 & 0.66 & -6.0
\end{pmatrix},
\\[6pt]
{\bf h}^{+-}
&=
\begin{pmatrix}
-7.4 & -1.9\\
-1.9 & -6.1
\end{pmatrix},
\\[6pt]
{\bf h}^{-+}
&=
\begin{pmatrix}
-7.5 & -1.8\\
-1.8 & -6.0
\end{pmatrix},
\\[6pt]
{\bf h}^{--}
&=
\begin{pmatrix}
-4.9
\end{pmatrix},
\end{align}

where all matrix elements are given in eV.

\section{One-body Hamiltonian of Fe$_2$O$_2$}
\label{app:fe2ham}

Here we provide the one-body matrices
${\bf h}^{\pi_x\pi_y\pi_z}$ of Eq. \eqref{eq:Fe2O2blocks}. In the orbital bases
defined by $\mathcal C_{\pi_x\pi_y\pi_z}$,
they read
\begin{align}
{\bf h}^{+++}
&=
\begin{pmatrix}
-4.639011 & 0.262112 & 1.007178\\
0.262112 & -5.601750 & -2.299061\\
1.007178 & -2.299061 & -8.333589
\end{pmatrix},
\\[6pt]
{\bf h}^{++-}
&=
\begin{pmatrix}
-5.101616 & -2.131874\\
-2.131874 & -7.778906
\end{pmatrix},
\\[6pt]
{\bf h}^{+-+}
&=
\begin{pmatrix}
-3.913771 & 2.784538\\
2.784538 & -6.569977
\end{pmatrix},
\\[6pt]
{\bf h}^{+--}
&=
\begin{pmatrix}
-4.665623 & 2.151210\\
2.151210 & -6.915002
\end{pmatrix},
\\[6pt]
{\bf h}^{-++}
&=
\begin{pmatrix}
-4.639011 & -0.262112 & -1.476990\\
-0.262112 & -4.727434 & -1.654394\\
-1.476990 & -1.654394 & -8.6786875
\end{pmatrix},
\\[6pt]
{\bf h}^{-+-}
&=
\begin{pmatrix}
-4.285882
\end{pmatrix},
\\[6pt]
{\bf h}^{--+}
&=
\begin{pmatrix}
-5.176413 & -2.838414\\
-2.838414 & -7.8898315
\end{pmatrix},
\\[6pt]
{\bf h}^{---}
&=
\begin{pmatrix}
-4.665623
\end{pmatrix},
\end{align}
where all matrix elements are given in eV.

\end{document}